\documentclass[aps, prd, amsmath, floats, floatfix, onecolumn, nofootinbib]{revtex4}
\pdfoutput=1
\usepackage{graphicx}
\usepackage{color}
\usepackage{latexsym}

\usepackage{amssymb}

\newcommand{\be}{\begin{eqnarray}}
\newcommand{\ee}{\end{eqnarray}}

\newcommand{\bea}{\begin{eqnarray}}
\newcommand{\eea}{\end{eqnarray}}

\begin{document}

\title{Cutkosky rules and perturbative unitarity \\
in Euclidean nonlocal quantum field theories}

\author{Fabio Briscese}\email{briscese.phys@gmail.com, briscesef@sustc.edu.com}

\affiliation{SUSTech Academy for Advanced Interdisciplinary Studies, Southern University of Science
and Technology, Shenzhen 518055, China}

\affiliation{Istituto Nazionale di Alta Matematica Francesco
	Severi, Gruppo Nazionale di Fisica Matematica, Citt\`{a}
	Universitaria, P.le A. Moro 5, 00185 Rome, Italy.}

\author{Leonardo Modesto}\email{lmodesto@sustc.edu.com}

\affiliation{Department of Physics, Southern University of Science
and Technology, Shenzhen 518055, China}

\begin{abstract}
We prove the unitarity of the Euclidean nonlocal scalar field theory to all perturbative orders in the loop expansion. The amplitudes in the Euclidean space   are calculated assuming that all the particles have purely imaginary energies, and afterwards  they are analytically continued  to real energies.  We show that such amplitudes satisfy the Cutkowsky rules and that only the cut diagrams corresponding to normal thresholds contribute to their imaginary part. This implies that the  theory is unitary. This analysis is  then exported to nonlocal gauge and gravity theories by means of Becchi-Rouet-Stora-Tyutin or diffeomorphism invariance, and Ward identities.

\end{abstract}

\maketitle

\section{INTRODUCTION} 
Nonlocal field theories are earning growing interest in the scientific community, since it has been realized that nonlocality might be a key ingredient  for the formulation of a quantum renormalizable theory of gravitation.  Nonlocal quantum gravity was proposed about 30 years ago by Krasnikov  \cite{Krasnikov} as a superrenormalizable theory for the gravitational interaction, and later extensively studied by Kuz'min (1989) \cite{kuzmin}.  
A very general class of local superrenormalizable theories were  proposed and extensively studied for the first time in \cite{shapiro asorey}.
However, only recently have nonlocal $D$-dimensional unitary and superrenormalizable theories  been proposed in \cite{modesto,Modesto:2012ys}.

%

Lately, it has been proved that the theory is finite at any perturbative order in the loop expansion \cite{Modesto:2017sdr,modestoLeslaw}. The theory can be formulated in Minkowski or in Euclidean signature, but we cannot pass from one to the other by  means of a Wick rotation because of the behavior of the nonlocal form factors at infinity in the complex plane. In fact,  the  form factors considered here are analytic and do not have poles at finite momenta $k=(k^0,\vec{k})$, in order to ensure the unitarity of the theory; hence, they must diverge at infinity in some direction on the  plane of complex energies $k^0$. Therefore,  integrals containing such form factors, such as those defining the complex scattering amplitudes, are different when performed along the real or the complex axis, because the integrals on the arcs in the first and fourth quadrant of the $k^0$ plane are nonzero. Indeed, even when the same action can be defined both in the Minkowski and Euclidean spaces, the two choices will correspond to two different theories.

In this paper we prove the unitarity of  nonlocal theories defined in the Euclidean space.  We calculate the complex amplitudes of scattering processes, assuming that all  energies $k^0_i$ and $p^0_i$ of the loop and external momenta, respectively, are purely imaginary; indeed, all the integrals in the scattering amplitudes are finite, due to the nonlocal form factors. Then, we reconstruct the amplitudes for real values of the external energies by a proper analytic continuation. Finally, we show that the imaginary part of the amplitudes constructed following this procedure is given by  Cutkosky rules. Moreover, only cut diagrams corresponding to normal thresholds contribute to such imaginary part, exactly as in the case of local field theories, so that  the unitarity of the theory is preserved.  In our proof, we do not make use of the reflection positivity in order to avoid any related issues
\cite{reflection positivity}.

We are interested in three main examples of nonlocal field theories: the scalar, gauge, and gravitational nonlocal theories introduced below. However,  for the sake of simplicity, we derive  Cutkosky rules in the case of a scalar field, and then we discuss how this derivation can be  straightforwardly extended to the gauge and gravitational cases. Nonlocal scalar field theory is described by the following nonlocal Lagrangian:
\be
\mathcal{L}_{\phi} = -\frac{1}{2}  \phi \, e^{H(- \sigma \Box)}\left(\Box +  m^2 \right)\phi  - \lambda \sum_{n=4}^{N} \frac{c_n}{n !} \phi^n \, ,
\label{phin1}
\ee
where $\exp H( -\sigma \Box)$ is the nonlocal form factor,  $\sigma$ is a parameter with dimensions $[\sigma] = - 2$ that fixes  the nonlocality scale $\ell_\Lambda = \sqrt{\sigma}$,  $N$ is an integer number, and $c_n$ are constant parameters.  The function $H(z)$ must be entire, i. e. ,  analytic with no poles at finite $z$, and it is assumed to be a polynomial or an asymptotically logarithmic function in the most interesting cases. The latter class of form factors is need for constructing consistent non-Abelian gauge theories or gravitational theories [see (\ref{Hkuzmin}) and (\ref{HTombo})]. However, in order to prove unitarity, we only need to assume that the form factor does not have zeros and poles at finite values of the loop four-momenta $k$, and that it goes to infinity for $k^0 \rightarrow \pm i \infty$ and $\vec{k} \rightarrow \pm  \infty$, ensuring the finiteness or renormalizability of the scattering amplitudes. Below we study a completely equivalent formulation of  the nonlocal scalar field theory, which is obtained from (\ref{phin1}) by the following field redefinition:
\bea\label{field redefinition}
\phi = e^{ - \frac{1}{2} H(- \sigma \Box) } \varphi,
\eea
which recasts (\ref{phin1}) in 
\be
\mathcal{L}_{\phi} = \frac{1}{2} \partial_\mu \varphi \, \partial^\mu \varphi - \frac{1}{2} m^2 \varphi^2 - \lambda \sum_{n=4}^{N} \frac{c_n}{n !} (  e^{ - \frac{1}{2} H(- \sigma \Box) } \varphi )^n \, .
\label{phin2}
\ee
The scalar field  propagator of (\ref{phin2}) is  the same as the one of a local two derivative theory, and it satisfies the Kallen-Lechmann representation \cite{peskin,itzykson zuber}, while nonlocality is restricted to the interaction terms.

Notice that the field redefinition (\ref{field redefinition}) implies a change in the path integral measure. However, such change has no effect on the quantum effective action and on the scattering amplitudes. Indeed, the particular field redefinition (\ref{field redefinition}) 
changes the integration over the field $\phi$ into an integration over the new field $\varphi$ in the functional integral introducing the Jacobian 
$||J||$ for the transformation 
\be
\prod_x \frac{\delta \phi(x)}{\delta \varphi(x)} = \prod_x \exp - \frac{1}{2} H(- \sigma \Box_x). 
\ee
The transformation (\ref{field redefinition}) is invertible because the form factor $\exp  - \frac{1}{2} H(z)$ is never zero for finite values of $z$; hence, one has $||J||\neq 0$. Moreover, the Jacobian $||J||$ is constant, because it does not depend on the fields; hence, it cancels in the calculation of the normalized generating functional $Z(j)/Z(0)$ (see \cite{kuzmin,antoniadis} for more details). In conclusion, the two theories (\ref{phin1}) and (\ref{phin2}) give the same scattering amplitudes at classical and quantum levels: hence, they are equivalent. 
%

Nonlocal gauge theory was proposed in 
\cite{kuzmin}
as a superrenormalizable one, but afterwards a generalization of such theory was shown to be finite in \cite{Modesto:2015lna,piva}. The action for the finite nonlocal Yang-Mills theory in flat spacetime reads
\be \!\! 
{S}_{\rm YM} = -\frac{1}{2 g_{\rm YM}^2} \int d^D x {\rm tr}\Big[ {\bf F}e^{H(\sigma {\cal D}^2)}
{\bf F} + s_g \, \sigma^2 \,  {\bf F}^2 \left( \sigma {\cal D}^2 \right)^{\gamma-2} {\bf F}^2 \Big]  \, . 
\label{gauge}
\ee
The notation on flat spacetime reads as follows: we use the gauge-covariant box operator defined via ${\cal D}^2_\Lambda={\cal D}_\mu{\cal D}^\mu$, where ${\cal D}_\mu$ is a gauge-covariant derivative (in the adjoint representation) acting on gauge-covariant field strength ${\bf F}_{\rho\sigma} = F_{\rho\sigma}^a T^a$ of the gauge potential $A_{\mu}$ (where $T^a$ are the generators of the gauge group in the adjoint representation.) Moreover, $\sigma = \ell_{\Lambda}^2 = 1/\Lambda^2$ ($[\sigma] =2$), where $\ell_\Lambda$ is the nonlocality length scale,  $s_g$ is a dimensionless parameter, and $\gamma \geqslant 2$ is an integer properly selected in order to have divergences only at one loop. Finally, two special examples of form factors are given in (\ref{Hkuzmin}) and (\ref{HTombo}). The same physical scale can be expressed in terms of a length scale $\ell_\Lambda$ or an energy scale $\Lambda$.

The action for nonlocal gravity \cite{kuzmin,
modesto,modestoLeslaw} consists in the Einstein-Hilbert term, a nonlocal operator that is quadratic in the Ricci tensor, and a potential at least cubic in the Riemann tensor. The action reads
\be
S_{\rm g}  =  - \frac{2}{\kappa^{2}_D}  \int  d^D x \sqrt{-g} \left[ R + G_{\mu\nu} \gamma(\Box) R^{\mu\nu} + V(\mathcal{R}) \right] \, .
\label{gravity}
\ee
where $R$, $ R_{\mu\nu}$, and $G_{\mu\nu}$  are the Ricci scalar, Ricci curvature, and the  Einstein tensor, respectively. Moreover, $V(\mathcal{R})$ is a generalized potential and $\mathcal{R}$ stands for scalar, Ricci, or Riemann curvatures, and derivatives thereof. The form factor $\gamma(\Box)$ depends on the nonlocality scale $\ell \equiv \sqrt{\sigma}$ and is defined by 
\be
\gamma(\Box) = \frac{e^{H(\sigma \Box)} -1}{\Box} \, .
\ee

Two examples of form factors $\exp H(z)$, which are suitable for gravity as well as for gauge theory, are
\be
&& \hspace{-0.8cm}
 H_{\rm K}(z) = \alpha \left[ \log (z )+\Gamma (0,z)+\gamma_E \right]  \, , 
 \,\,\,   {\rm Re} \, z > 0 \, ,   
  \label{Hkuzmin}\\
 &&\hspace{-0.8cm}
H_{\rm T}(p) = \frac{\alpha}{2} \left[ \log \left(p^2\right)+\Gamma \left(0,p^2\right)+\gamma_E \right] \, , \,\,\,   {\rm Re} \, p^2 > 0 \, ,
\label{HTombo}
\ee
where  $\gamma_E \approx 0.577216$ is the Euler-Mascheroni constant,  
$
\Gamma(0,x) = \int_x^{+ \infty}  d t \, e^{-t} /t 
$, $p\equiv p(z)$ is a polynomial of $z = - \sigma \Box$ of degree $\gamma+1$ with $\gamma > D/2$, 
and $\alpha$ is an integer number. Both the entire functions (\ref{Hkuzmin}) and (\ref{HTombo}) are asymptotically polynomial in a conical region around the real axis, as required by the locality of the counterterms \cite{modestoLeslaw}. 
Exponential form factors like $\exp[ (-\Box)^n]$ are suitable for scalar theories, but not for gravity or gauge theories.

The propagators of the gauge and  graviton fields are modified by the form factors, yet without introducing any extra poles besides the gauge fields and the graviton. Therefore, in momentum space ($k_\mu$) 
the propagators for the theories (\ref{gauge}) and (\ref{gravity}), respectively, read
\be
&&
G(k)^{\rm YM}_{\mu\nu \, ab } = - i  \delta_{ab} \frac{ e^{-H(\sigma k^2) }}{k^2 +  i \epsilon } \left( \eta_{\mu\nu} -     \frac{k_\mu k_\nu}{k^2} \right) + {(-   \delta_{ab})} i \xi_{\rm ym} \frac{k_\mu k_\nu}{\omega_{\rm ym}(k^2) (k^2 { + i \epsilon})}
 \, , 
\label{NLPym}
\\
&& 
G(k)^{\rm g} = i \frac{ e^{-H(\sigma k^2) }}{k^2 + i \epsilon } \left( P^{(2)} - \frac{1}{D-2} P^{(0)} \right)
+ 
i  \frac{\xi_{\rm g} (2P^{(1)} + \bar{P}^{(0)} ) }{2 (k^2  {+ i \epsilon}) \, \omega_{\rm g}( k^2/\Lambda^2)}  \, ,
\label{NLPg}
\ee
where $\xi_{\rm ym}$, $\xi_{\rm g}$ are gauge parameters and $\omega_{\rm ym}(k^2)$, $\omega_{\rm g}(k^2)$ are gauge-fixing weighting functions \cite{modesto}. 
The tensors $P^{(2)}$, $P^{(0)}$,  $P^{(1)}$, and $\bar{P}^{(0)}$ with four indices (here omitted) are the usual spin-projector operators \cite{modesto}.
The vertices in  theories (\ref{gauge}) and (\ref{gravity}) are very complicated, but not crucial to proving unitarity. For this purpose, it is sufficient to assume that they are analytic functions of the momentum. 
Moreover, we assume they come from a gauge-invariant  or Becchi-Rouet-Stora-Tyutin (BRST)-invariant action.  
These last two assumptions are surely verified for our theories  (\ref{gauge}) and (\ref{gravity}). Therefore, it is straightforward to derive their unitarity once it is proved for the nonlocal scalar field theory, which does not involve complicated tensorial structure and gauge symmetry.

This paper is organized as follows: in Sec. \ref{Section Cutkosky rules} we define the complex amplitudes  for the Euclidean nonlocal scalar theory  (\ref{phin2}) and derive the  Cutkosky rules. In Sec. \ref{Section unitarity and Cutkosky rules} we summarize the relation between unitarity and Cutkosky rules, discussing the issues related to anomalous thresholds. As an application, in Secs \ref{Section one loop} and \ref{Section two loops} we explicitly consider the cases of one- and two-loop diagrams with no anomalous thresholds, proving their unitarity. In Sec. \ref{section unitarity} we complete the proof of the unitarity of the nonlocal scalar theory, showing that anomalous thresholds do not contribute to the imaginary part of the amplitudes. As an example, in Secs \ref{section triangle} and \ref{section Box} we prove the unitarity of the triangle and box diagrams, which do have anomalous thresholds.  Then, in Sec. \ref{section gauge gravity} we discuss  how these results are straightforwardly extended to the case of Euclidean nonlocal gauge and gravitational theories, on the basis of BRST invariance and Ward identities. Finally, in Sec. \ref{section conclusions} we  resume these results giving final remarks and conclude.

\section{Complex amplitudes and Cutkosky rules for Euclidean nonlocal scalar field theory} \label{Section Cutkosky rules}


In this section, we will define the complex amplitudes for the nonlocal scalar field (\ref{phin2}) in the Euclidean space, and we will prove that  their imaginary part is given by the Cutkosky rules, which are generalized to the class of theories under consideration.

Since our theory is defined in the Euclidean space, when
evaluating the complex amplitudes, the loop integrals are
calculated integrating the loop spatial moments in $\mathbb{R}^3$
and the loop energies along the imaginary axis $\mathcal{I}$. At the
first step, when performing the integrations, also the external
energies are taken to be purely imaginary. In this situation, all
the poles of the propagators are far from the integration region
$\mathcal{I} \times \mathbb{R}^3$ in each loop variable, and we
can safely integrate, obtaining a finite expression for the
amplitudes for purely imaginary energies. Subsequently, these
amplitudes are analytically continued to real and positive
external energies. Finally, it is  proved that the imaginary part of the
amplitudes obtained in this way is given by  the Cutkosky rules. 

\begin{figure}
	\begin{center}
		\includegraphics[height=10cm]{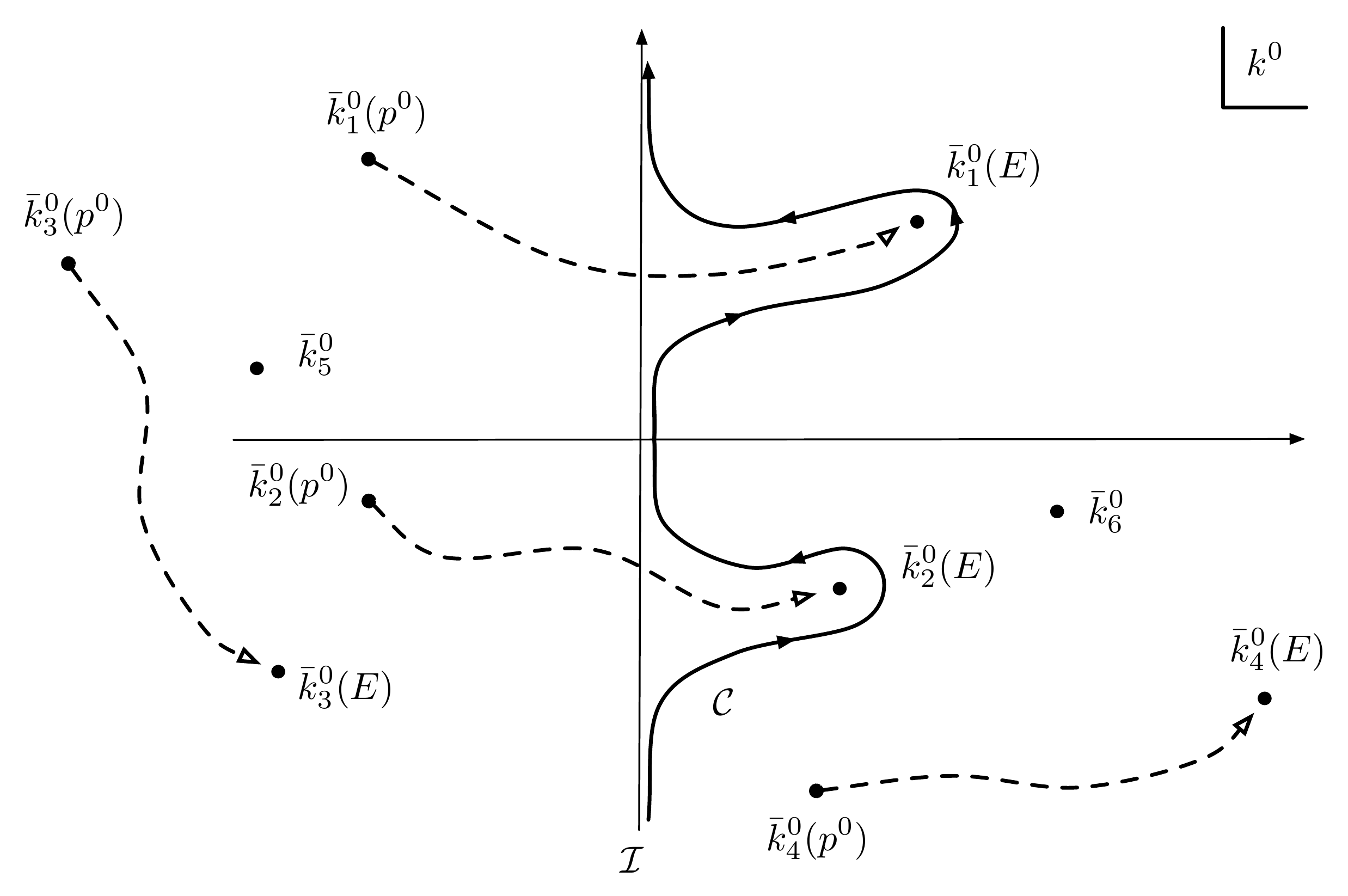}
		\caption{We schematically plot the poles of the propagators on the $k^0$ complex plane, where $k$ is one of the loop momenta. We see that, when the external energy $p^0$ is sent to its real and positive value $E$, some pole moves in the $k^0$ plane. Some of these moving poles  pass through the imaginary axis $\mathcal{I}$, so that the integration contour $\mathcal{C}$ is obtained deforming $\mathcal{I}$ around them.}
		\label{Fig4}	
	\end{center}
\end{figure}

For obvious practical limits, the  analytic continuation is not obtained from the explicit integrated expressions of the amplitudes. In fact, such explicit expressions would depend on the specific form of the nonlocal vertices.  Instead, we continue analytically  the generic integral expressions of the amplitudes, considering the limit in which the external energies pass from purely
imaginary to real positive values. Indeed, our results will be valid for any nonlocal form factor.  However, we must pay special attention in performing such limit, since when we move the external energies from purely imaginary  to real positive values, some of the poles of the propagators move through the imaginary axes $\mathcal{I}$. Therefore, the analytic continuation of the amplitudes is obtained deforming  $\mathcal{I}$ around these moving poles. This situation is schematically represented in Fig.\ref{Fig4}. As we will see, for some values of the external energies the new integration contour $\mathcal{C}$  is pinched by the poles and cannot be deformed further, so that the complex amplitudes become singular.

Let us be more specific, and let us start considering  a general scattering amplitude for a process with $L$-loop in the theory (\ref{phin1}).  For a generic diagram with $I$ internal lines and $V$ vertices, the scattering amplitude will be
\begin{equation}\label{amplitude1}
i  \mathcal{M} = \frac{1}{S_{\#}} \int_{(\mathcal{I}\times
\mathbb{R}^3)^I} \prod_{i=1}^I \frac{d^4 k_i}{(2\pi)^4}
\frac{i}{k_i^2 - m^2 + i \epsilon} \prod_{j=1}^{V} (- i)  \lambda
\mathcal{V}(p^{(j)} ) (2 \pi)^4 
\delta^{(4)} \left(\sum_{\ell = 1}^{ N}  p^{(j)}_\ell \right)
\, ,
\end{equation}   
where $\mathcal{I}$ is the imaginary axis, $S_{\#}$ is a combinatoric factor, and the $\delta^{(4)} \big( \sum_\ell  p^{(j)}_\ell \big)$ gives the momentum conservation at the $j$-th vertex.  The vertex terms $\mathcal{V}(p^{(j)}_\ell )$ come from the nonlocal interaction at the $j$-th vertex. If we consider a polynomial interaction $\lambda \, \phi^N/N!$ in (\ref{phin1}), corresponding to an interaction term $\lambda\,(\exp[ - H(- \sigma \Box)/2 ] \varphi)^N/N!$  in (\ref{phin2}), the vertex function $\mathcal{V}(p^{(j)}_\ell )$ will be
\begin{equation}\label{vertex V definition}
\mathcal{V}(p^{(j)} ) = \prod_{i=1}^{N} e^{ - \frac{ H\left(\sigma \left(p_{i}^{(j)} \right)^2\right)}{2}  },
\end{equation}
where $p^{(j)}_1, \ldots p^{(j)}_N$ are the momenta at the $j$-th vertex. For simplicity, hereafter we will consider vertex functions of the form (\ref{vertex V definition}). However, the results presented below are valid for any interaction term of the type given in (\ref{phin1}) and (\ref{phin2}).

After integration of the delta functions in (\ref{amplitude1}) one has
\begin{equation}\label{amplitude2}
\mathcal{M} = - \frac{ \lambda^V  }{S_{\#} }\,
\int_{(\mathcal{I}\times \mathbb{R}^3)^L} \prod_{i=1}^L \frac{i \,
d^4 k_i}{(2\pi)^4} \frac{1}{k_i^2 - m^2 + i \epsilon}
\prod_{j=1}^{I-L} \frac{1}{q_j^2 - m^2 + i \epsilon} \,
B(k_i,p_h)\, ,
\end{equation}
where $q_j$ is a linear combination with coefficients $\pm 1$ of
the loop momenta $k_i$ and of the external momenta $p_h$. In (\ref{amplitude2}) we have
used the topological relation $I-V = L-1$, which implies $i^{I-V}= i^{L-1}$, and we have defined the function $B(k_i,p_h)$ as
\bea\label{generic B}
B(k_i,p_h)\equiv \prod_{j=1}^{V} \mathcal{V}(p^{(j)}_\ell),
\eea

We note that the nonlocality of the theory is encoded only in the term $B(k_i,p_h)$, which ensures the ultraviolet convergence of the integrals in the scattering amplitudes. Since by hypothesis  $B(k_i,p_h)$ has no zeros or poles in the complex hyperplane $(\mathbb{C}\times
\mathbb{R}^3)^{L+N}$ for finite momenta $k_i$ and $p_h$, this function does not introduce other poles in the integrand in (\ref{amplitude2})  than those of the propagators. Indeed, the nonlocality  does not change the singularity structure of the scattering amplitude with respect to the case of a local scalar field, and it does not play any role in what we will discuss below.

As explained above, when evaluating (\ref{amplitude2}) we first take purely imaginary values of the external energies $p^0_h$, so that the poles of the propagators are far from the integration contour. In fact, $k_i^2 - m^2 \neq 0 $ and $q_j^2 - m^2 \neq 0$ when $k^0_i$ and $p^0_h$ are purely imaginary. Afterwards, we consider the analytic continuation of   (\ref{amplitude2}), taking the limit in which the external energies go to their physical real values, i.e., $p^0_h \rightarrow E_h \in \mathbb{R}^+_0$. This is obtained
deforming the integration contour $\mathcal{I}\times \mathbb{R}^3$ for each loop around the poles that pass through the imaginary  axis $\mathcal{I}$ in the limit procedure, but keeping the integration end points to $\pm i \infty$ in order to guarantee the ultraviolet convergence of the loop integrals.\footnote{In gravity and gauge theories the convergence is guaranteed by the introduction of other local or nonlocal operators in the action.} 

Therefore, the analytic continuation of the amplitude (\ref{amplitude2}) will be given by
\begin{equation}\label{amplitude2bis}
\mathcal{M} = - \frac{ \lambda^V  }{S_{\#} }\,
\int_{(\mathcal{C}_L\times \mathbb{R}^3)}
\int_{(\mathcal{C}_{L-1}\times \mathbb{R}^3)} \ldots
\int_{(\mathcal{C}_1\times \mathbb{R}^3)}\, \prod_{i=1}^L \frac{i
\, d^4 k_i}{(2\pi)^4} \frac{1}{k_i^2 - m^2 + i \epsilon}
\prod_{j=1}^{I-L} \frac{1}{q_j^2 - m^2 + i \epsilon} \,
B(k_i,p_h)\, ,
\end{equation}
where  $\mathcal{C}_i\times \mathbb{R}^3$ is the deformed contour
for the $i$-th integration variable.

The deformation of the integration contour around the poles piercing the imaginary axis is allowed when such contour is not constrained between separate poles that overlap in the limit procedure in which the external energies go to their real values, and this is always possible whether the Feynman $i \epsilon$ prescription is implemented. On the contrary, when $\epsilon \rightarrow 0$ the integration contour will be constrained between poles that are initially separate but coincide for some real value of the external energies. When this happens, the integration contour cannot be deformed further, and this implies that the amplitude has a singularity at such energies, and  develops a branch cut discontinuity. Therefore, the amplitude (\ref{amplitude2bis}) will have poles and branch cuts at at some values of the external energies in the limit $\epsilon \rightarrow 0$.

In the case of local fields, the singularities of the amplitudes are obtained from the Landau equations \cite{landau},  which simply express the conditions that two  of the propagator poles coincide at some values of the external momenta, so that the integrand in (\ref{amplitude2bis}) has a double pole; see \cite{itzykson zuber} for a review of the Landau equations. However, since the poles of the local and nonlocal theories are the same, the singularities of the local and nonlocal theories will be the same. Furthermore, the singularities of the amplitudes in nonlocal theories will be determined by the same Landau equations \cite{landau} valid for local field theories. Therefore, the nonlocality does not change  the positions of the poles of the propagators or the singularity structure of the diagrams.

Having established this fact, we have  to show that  the imaginary part of the complex amplitudes at the branch cuts corresponding to the Landau poles is given by the Cutkosky rules, as in the local case.  Therefore, we have to calculate the limit  
\be
\lim_{\epsilon \rightarrow 0} \left\{\mathcal{M}(E_h, \epsilon) - \mathcal{M}^*(E_h, \epsilon)\right\}  \, ,
\label{Limit M}
\ee 
for the amplitude (\ref{amplitude2bis}).  Since in the original derivation of the Cutkosky rules in \cite{cutkosky}, Equ. (\ref{Limit M}) is expressed as the discontinuity of $\mathcal{M}$ at the branch cut, we first prove that, also in the nonlocal case, one has
\be
\lim_{\epsilon \rightarrow 0} \left\{\mathcal{M}(E_h, \epsilon) - \mathcal{M}^*(E_h, \epsilon)\right\} = \lim_{\epsilon \rightarrow 0} \left\{\mathcal{M}(E_h,\epsilon) - \mathcal{M}(E_h,- \epsilon)\right\}
\label{Limit M2}  \, .
\ee 
Let us consider a specific amplitude $\mathcal{M}_{ba} = \langle b | \mathcal{M} | a \rangle$ obtained by means of the analytic continuation described before as in (\ref{amplitude2}), that is

\begin{equation}\label{Mba} 
\mathcal{M}_{ba}(p_h,\epsilon) = - \frac{ \lambda^V  }{S_{\#} }\,
\prod_{i=1}^L \int_{- i \infty \, (\mathcal{C})}^{+ i \infty} \int_{\mathbb{R}^3} \frac{i \,
	dk^0_i d^{3} k_i}{(2\pi)^4} \frac{1}{k_i^2 - m^2 + i \epsilon}
\prod_{j=1}^{I-L} \frac{1}{q_j^2 - m^2 + i \epsilon} \,
B( k_i, p_h)\, ,
\end{equation}
where we have explicitly indicated the dependence from the external energies $p_h$ and the parameter $\epsilon$. 
The notation used also  indicates that the integration contour $\mathcal{C}$ is oriented from $- i \infty$ to $+ i \infty $. The amplitude corresponding to the inverse process $b \rightarrow a$ is given by the same expression as in (\ref{Mba}), i.e., $\mathcal{M}_{ab} = \mathcal{M}_{ba}$. 

The complex conjugate of the amplitude is
\be
\label{Mab conj1} 
&& \mathcal{M}_{ab}^*(p_h,\epsilon) = - \frac{ \lambda^V  }{S_{\#} }\left[
\prod_{i=1}^L \int_{- i \infty \, (\mathcal{C})}^{+ i \infty} \int_{\mathbb{R}^3} \frac{i \,
	dk^0_i d^{3} k_i}{(2\pi)^4} \frac{1}{k_i^2 - m^2 + i \epsilon}
\prod_{j=1}^{I-L} \frac{1}{q_j^2 - m^2 + i \epsilon} \,
B( k_i, p_h) \right]^* \nonumber	\\
&& \hspace{1.8cm} = - \frac{ \lambda^V  }{S_{\#} }\,\prod_{i=1}^L \int_{+ i \infty \, (\mathcal{C}^*)}^{- i \infty} \int_{\mathbb{R}^3} \frac{- i \,
	d k^{0 *}_i d^{3} k_i}{(2\pi)^4} \frac{1}{(k^*_i)^2 - m^2 - i \epsilon}
\prod_{j=1}^{I-L} \frac{1}{(q^*_j)^2 - m^2 - i \epsilon} \,
B( k^*_i, p^*_h)\, .
\ee
Redefining the name of the integration variable $k^*_i \rightarrow k_i$ one has
\be\label{Mab conj2} 
\mathcal{M}_{ab}^*(p_h,\epsilon) =- \frac{ \lambda^V  }{S_{\#} }\,\prod_{i=1}^L \int_{+ i \infty \, (\mathcal{C}^*)}^{- i \infty} \int_{\mathbb{R}^3} \frac{- i \,
	dk^0_i d^{3} k_i}{(2\pi)^4} \frac{1}{k_i^2 - m^2 - i \epsilon}
\prod_{j=1}^{I-L} \frac{1}{\tilde{q}_j^2 - m^2 - i \epsilon} \,
B( k_i, p^*_h)\, ,
\ee
where the four vector $\tilde{q}_i$ are obtained by the replacement $p^0 \rightarrow (p^0)^*$ in $q_i$. The integration contour $\mathcal{C}^*$ is obtained from $\mathcal{C}$ by reflection with respect to the real axis, 
and it is orientated from $+ i \infty$ to $-i\infty$. 
Moreover, the complex conjugation changed the Feynman $i \epsilon$ prescription in $-i \epsilon$. 

When evaluated for real and positive external energies $p^0_h = E_h$, (\ref{Mab conj2}) gives
\be\label{Mab conj3} 
\mathcal{M}_{ab}^*(E_h,\epsilon)  =- \frac{ \lambda^V  }{S_{\#} }\,\prod_{i=1}^L \int_{- i \infty \, (\mathcal{C}^*)}^{+ i \infty} \int_{\mathbb{R}^3} \frac{i \,	dk^0_i d^{3} k_i}{(2\pi)^4} \frac{1}{k_i^2 - m^2 - i \epsilon}
\prod_{j=1}^{I-L} \frac{1}{q_j^2 - m^2 - i \epsilon} \,
B( k_i, p_h)\,  ,
\ee
where we have inverted the orientation of $\mathcal{C}^*$ reabsorbing a minus sign, and used the fact that $\tilde{q}_i = q_i$ for real $p^0_h$.
Therefore, Eq. (\ref{Limit M2}) follows straightforwardly from (\ref{Mab conj3}).

At that point we can show that the discontinuity of $\mathcal{M}$ in the rhs. of (\ref{Limit M2}) is given by the Cutkosky rules. Let us consider the case of two poles that converge in the limit $\epsilon \rightarrow 0$ at some real external energies $p^0_h = E_h$ corresponding{\tiny } to a specific Landau singularity. In Fig.\ref{Fig4}, this situation is represented by the pole $\bar{k}^0_2(E)$, which approaches $\bar{k}^0_6$ and constrains the integration contour $\mathcal{C}$. Since (\ref{amplitude2bis}) is of the same form of the integrals considered in \cite{cutkosky}, we can make use of the result of Cutkosky (skipping any useless rederivation of such result), which states that the discontinuity of $\mathcal{M}$ at the branch cut is given by the residue of the integrand in  (\ref{amplitude2bis}) at the pole, times a $(2 \pi i)$ numerical factor. We stress that only the propagators that go on shell at the Landau singularity under consideration contribute to (\ref{Limit M}). Therefore,  (\ref{Limit M2}) is obtained replacing in  (\ref{amplitude2bis}) each propagator that goes on shell at the singularity with a term  $(- 2 \pi i) \,\delta(p^2-m^2)$, exactly as in the case of a local scalar field theory. The only difference with respect to the local case is that one has to  replace the integration region $\mathcal{C}\times \mathbb{R}^3$ with $\mathbb{R}^4$ for all the loop momenta contained in the on shell propagators. Indeed, the Cutkosky rules are still valid, \textit{mutatis mutandis}, for the class of nonlocal theories under consideration.

\begin{figure}
	\begin{center}\label{figurepoles}
		\includegraphics[height=8cm]{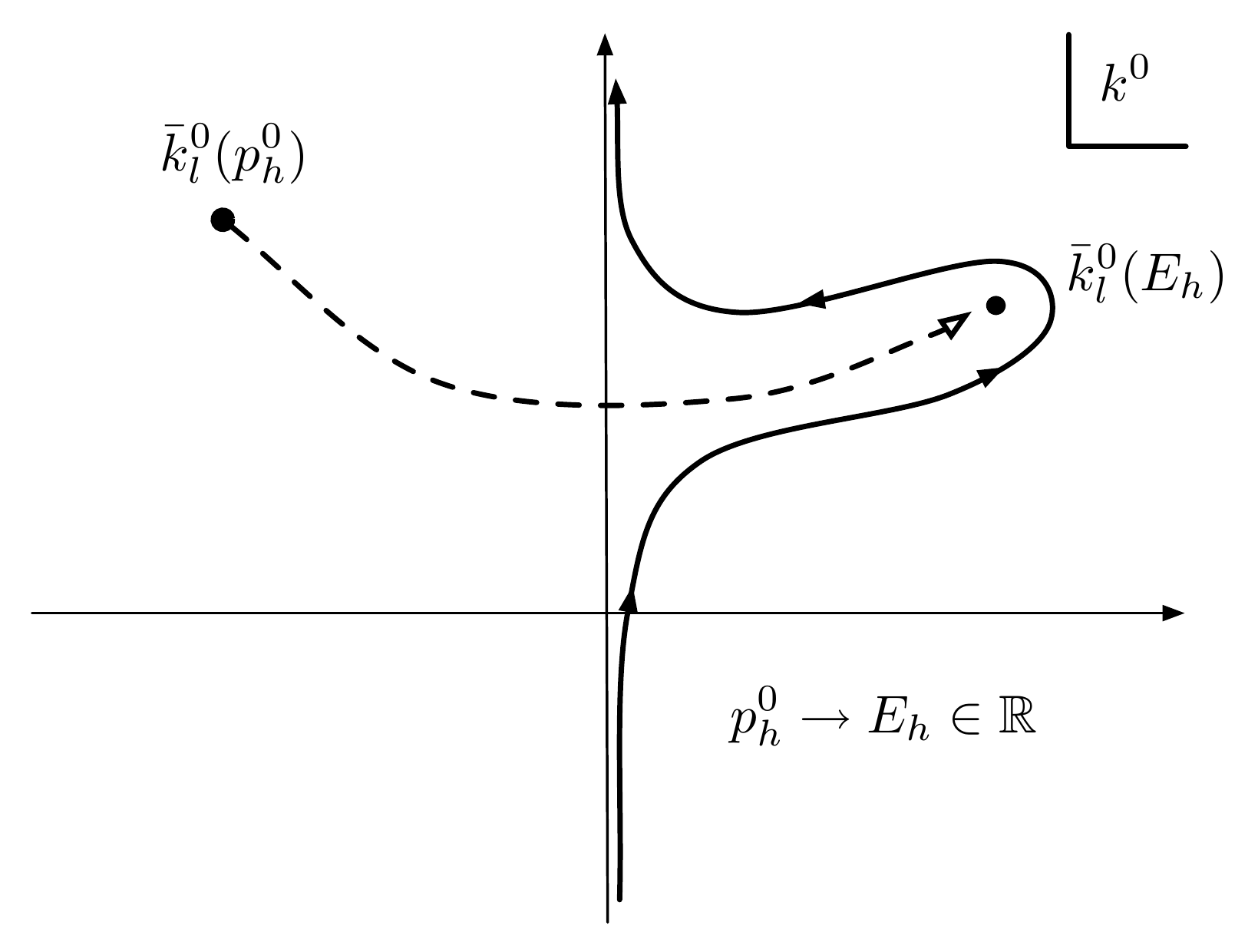}
		\caption{The pole $\bar{k}^0_l(p_h)$ moves thorough the imaginary axes $\mathcal{I}$ when $p^0_h \rightarrow E^0_h$, and the integration contour $\mathcal{C}$ is obtained deforming  $\mathcal{I}$ around $\bar{k}^0_l(E_h)$.}
	\end{center}
\end{figure}

We can make the derivation of the Cutkosky rules given above even more explicit, giving a procedure to evaluate the analytically continued amplitude  (\ref{amplitude2bis}) for generic diagrams. In order to proceed, we must determine how the integration contour
$(\mathcal{I}\times \mathbb{R}^3)$ for the $i$-th loop variable
$k_i$ is deformed into $(\mathcal{C}_i\times \mathbb{R}^3)$. In what follows, we will give a prescription for finding $(\mathcal{C}_i\times \mathbb{R}^3)$ proceeding iteratively. Since we will integrate (\ref{amplitude2bis}) one loop at time, it is useful to define how the deformed integration contour is obtained for a generic integral in one-loop variable. Let us consider the following integral:
\begin{equation}\label{formula integral 0}
M(p_h) = \int_{\mathcal{I}\times \mathbb{R}^3} f(k,p_h) dk^0 d^3k \, ,
\end{equation}
where $p_h$ represents the external momenta, and the function
$f(k,p_h)$ has $m$ poles $\bar{k}^0_l(\vec{k},p_h)$ with $l=1,2,\ldots m$ in the complex
$k^0$ plane. We make the hypothesis that the poles
$\bar{k}^0_l(\vec{k},p_h)$ are not purely imaginary when the
external energies $p^0_h$ are purely imaginary. Indeed, $\bar{k}^0_l(\vec{k},p_h)$ are far from the integration contour for purely imaginary external energies, which is exactly the case of (\ref{amplitude2}). When we move
$p^0_h$  to real positive values, the poles $\bar{k}^0_l(\vec{k},p_h)$ will
move in the $k^0$ complex plane, and some of them will pass
through the integration contour $\mathcal{I}$ (see Fig.\ref{figurepoles}). Therefore, the analytic continuation of the function $M(p_h)$ to real values of
the external energies $p_h$ is obtained from (\ref{formula
integral 0}) deforming the integration contour  $\mathcal{I}$ around the
poles that pinch the imaginary axis, i.e.,

\begin{equation}\label{formula integral bis}
M(p_h) = \int_{\mathcal{C}\times \mathbb{R}^3} f(k,p_h) dk^0 d^3k
\, .
\end{equation}
If the integrand $f(k,p_h)$ is analytic, which will be always
assumed hereafter, the integral on $\mathcal{C}$ equals the
integral along the imaginary axis $\mathcal{I}$ plus the
contributions of the residues of $f(k,p_h)$ at each pole that has
passed through the imaginary axes in the limit $p^0_h \rightarrow
E_h \in \mathbb{R}^+_0$, that is,

\begin{equation}\label{formula integral 1}
M(p_h) = \int_{\mathcal{I}\times \mathbb{R}^3} f(k,p_h) \, dk^0
d^3k \pm (2 \pi  i) \int_{\mathbb{R}^3} \sum_l
{\rm Res}\left\{f(k,p_h), \bar{k}^0_l(\vec{k},p_h) \right\} \,
d^3k \,.
\end{equation}
where the sum on the index $l$ in the second
integral in (\ref{formula integral 1}) is limited to
those poles that passed through the imaginary axes. The plus sign  in (\ref{formula integral 1}) corresponds to poles that pass through $\mathcal{I}$ from left to right, while the minus sign corresponds to poles that move from right to left.

In order to evaluate the complex amplitude (\ref{amplitude2}), we can write $\mathcal{M}$ as

\begin{equation}\label{amplitude3}
\mathcal{M} = - \frac{ \lambda^V  }{S_{\#} }\,
\int_{(\mathcal{C}_L\times \mathbb{R}^3)} \frac{i \, d^4
k_L}{(2\pi)^4} \frac{F_L(k_L,p_h)}{k_L^2 - m^2 + i \epsilon}  \, ,
\end{equation}
where we have defined
\begin{equation}\label{definition FL 1}
F_L(k_L,p_h) = \int_{(\mathcal{C}_{L-1}\times \mathbb{R}^3)}
\int_{(\mathcal{C}_{L-2}\times \mathbb{R}^3)} \ldots
\int_{(\mathcal{C}_1\times \mathbb{R}^3)}\, \prod_{i=1}^{L-1}
\frac{i \, d^4 k_i}{(2\pi)^4} \frac{1}{k_i^2 - m^2 + i \epsilon}
\prod_{j=1}^{I-L} \frac{1}{q_j^2 - m^2 + i \epsilon} \,
B(k_i,p_h)\, ,
\end{equation}
and proceeding iteratively, we define
\begin{equation}\label{definition FL 2}
F_L(k_L,p_h) = \int_{(\mathcal{C}_{L-1}\times \mathbb{R}^3)}
\frac{i \, d^4 k_{L-1}}{(2\pi)^4}
\frac{F_{L-1}(k_L,k_{L-1},p_h)}{k_{L-1}^2 - m^2 + i \epsilon} \, ,
\end{equation}
with
\begin{equation}\label{definition FL 3}
F_{L-1}(k_L,k_{L-1},p_h) = \int_{(\mathcal{C}_{L-2}\times
\mathbb{R}^3)}  \frac{i \, d^4 k_{L-2}}{(2\pi)^4}
\frac{F_{L-2}(k_L,k_{L-1},k_{L-2},p_h)}{k_{L-2}^2 - m^2 + i
\epsilon} \, ,
\end{equation}
and so on, until the last expressions for $F_2$ and $F_1$, namely 
\bea 
\label{definition FL1 4}
&& F_{2}(k_L,k_{L-1},\ldots,k_{2},p_h) = \int_{(\mathcal{C}_1\times
\mathbb{R}^3)} \frac{i \, d^4 k_1}{(2\pi)^4}
\frac{F_{1}(k_L,k_{L-1},\ldots,k_{1},p_h)}{k_1^2 - m^2 + i
\epsilon}\, , \\
&&
\label{definition FL1 5}
F_{1}(k_L,k_{L-1},\ldots,k_{1},p_h) =  \prod_{j=1}^{I-L}
\frac{1}{q_j^2 - m^2 + i \epsilon} \, B(k_i,p_h)\, .
\eea
%
Equations (\ref{definition FL 1})-(\ref{definition FL 4}) can be recast by means of the recursive relation
\begin{equation}\label{definition FL 4}
F_{i+1}(k_L,k_{L-1},\ldots,k_{i+1},p_h) =
\int_{(\mathcal{C}_{i}\times \mathbb{R}^3)}  \frac{i \, d^4
k_{i}}{(2\pi)^4}
\frac{F_{i}(k_L,k_{L-1},\ldots,k_{i},p_h)}{k_{i}^2 - m^2 + i
\epsilon} \, .
\end{equation}

Therefore, the complex amplitude $\mathcal{M}$ can be obtained  evaluating the functions $F_{i}(k_L,\ldots,k_{i},p_h)$  by means of the integration formula (\ref{formula integral 1}) starting from
$F_{2}(k_L,k_{L-1},\ldots,k_{2})$  and then proceeding  iteratively, i.e.,

\begin{equation}\label{amplitude3 integrated}
F_{i+1}(k_L,\ldots,k_{i+1},p_h) =
\int_{(\mathcal{I}\times \mathbb{R}^3)} \frac{i \, d^4
	k_i}{(2\pi)^4} \frac{F_{i}(k_L,\ldots,k_{i},p_h)}{k_{i}^2 - m^2 + i
	\epsilon} \pm
\int_{(\mathbb{R}^3)} \frac{i\, d^3 k_i}{(2\pi)^4} \, (2\pi i)
\,\sum_l \, {\rm Res} \left\{\frac{F_{i}(k_L,\ldots,k_{i},p_h)}{k_{i}^2 - m^2 + i
	\epsilon},\bar{k}^0_{i,l}\right\} \, ,
\end{equation}

\begin{equation}\label{amplitude3 integrated2}
\mathcal{M} = - \frac{ \lambda^V  }{S_{\#} }\left[
\int_{(\mathcal{I}\times \mathbb{R}^3)} \frac{i \, d^4
k_L}{(2\pi)^4} \frac{F_L(k_L,p_h)}{k_L^2 - m^2 + i \epsilon} \pm
\int_{(\mathbb{R}^3)} \frac{i\, d^3 k_L}{(2\pi)^4} \, (2\pi i)
\,\sum_l \, {\rm Res} \left\{\frac{F_L(k_L,p_h)}{k_L^2 - m^2 + i
\epsilon},\bar{k}^0_{L,l}\right\} \right] ,
\end{equation}
where $\bar{k}^0_{i,l}\equiv \bar{k}^0_{i,l}(\vec{k_i},p_h)$ are the poles of $F_i(k_i, p_h)$ in
the $k^0_i$ complex plane, and the sum in $l$ is extended only to those
poles passing through $\mathcal{I}$ in the limit $p^0_h
\rightarrow E_h \in \mathbb{R}^+_0$.

We note that, in order to evaluate $F_{i+1}(k_L,k_{L-1}, \ldots,
k_{i+1},p_h)$ applying (\ref{formula integral 1}), one must find
the deformed contour $\mathcal{C}_i \times \mathbb{R}^3$ for the
$i$-th loop integration variable. Indeed, one has to study the poles of
the  function $F_{i}(k_L,k_{L-1}, \ldots, k_{i},p_h)$ and
determine which ones pass through the imaginary axes of the
complex plane of $k_i^0$ when we take the limit of real external
energies, \footnote{Note that the roots of $k_{i}^2 - m^2 + i
\epsilon =0$ do not depend on $p_h$ an do not pinch the imaginary
axis.} and this analysis must be repeated variable by
variable. For this purpose it is useful to stress that, when the $i \epsilon$ factor is maintained,  the poles of $F_{1}(k_L,k_{L-1}, \ldots,
k_{1},p_h)$ with positive imaginary part are well separated from those with negative imaginary part. Indeed, the deformed integration contour $\mathcal{C}_i \times \mathbb{R}^3$ cannot be pinched by the poles, and  $\mathcal{M}$ is not singular. Therefore, the singularities and the
corresponding branch cuts in $\mathcal{M}$ emerge in the limit $i \epsilon \rightarrow
0$.

Finally, having evaluated the complex amplitude by means of (\ref{amplitude3 integrated2}), the Cutkosky rules can be
obtained  evaluating
$\mathcal{M} - \mathcal{M}^*$ in the limit $\epsilon\rightarrow 0$. Again, here we do not expect any substantial difference with
respect to the case of a local field theory, since the propagators of local and
nonlocal theories have the same poles.

We stress that this derivation of the Cutkosky rules is useful to enlighten  the grounds of unitarity in nonlocal theories, but it is not necessary, since we have already shown that the amplitude (\ref{amplitude2bis}) falls under the hypothesis of the Cutkosky theorem. Indeed, the discontinuity in its imaginary part is obtained applying the result of Cutkosky.

Before concluding this section, it is appropriate to  comment briefly about the Lorentz invariance of the amplitude (\ref{amplitude2bis}), since this point might be questioned by the reader. Let us consider an infinitesimal Lorentz transformation $\Lambda$. Applying this transformation to external and internal momenta we have 

\begin{equation}\label{Lorentz 1}
\mathcal{M}^\prime = - \frac{ \lambda^V  }{S_{\#} }\,
\int_{(\mathcal{C}^\prime_L\times \mathbb{R}^3)}
\int_{(\mathcal{C}^\prime_{L-1}\times \mathbb{R}^3)} \ldots
\int_{(\mathcal{C}^\prime_1\times \mathbb{R}^3)}\, \prod_{i=1}^L \frac{i
	\, d^4 k_i}{(2\pi)^4} \frac{1}{k_i^2 - m^2 + i \epsilon}
\prod_{j=1}^{I-L} \frac{1}{q_j^2 - m^2 + i \epsilon} \,
B(k_i,p_h)\, ,
\end{equation}
where we have used the fact that $B(k_i,p_h)$ depends on $k_i$ and $p_h$ through the square of their linear combinations, i.e., $k_i^2$, $p_h^2$  and $q_i^2$. Therefore, $B(k_i,p_h)$ is Lorentz invariant. Moreover, the contours $\mathcal{C}^\prime_i$ are obtained from $\mathcal{C}_i$ through the Lorentz transformation $\Lambda$. Since $\Lambda$ is infinitesimal, the two contours $\mathcal{C}^\prime_i$ and $\mathcal{C}$ will be infinitesimally close, and we can safely assume that there will be no poles among them. Indeed, since the integrand in (\ref{Lorentz 1}) is analytic, we can replace the integration contours $\mathcal{C}^\prime_i$ with  $\mathcal{C}$, obtaining $\mathcal{M}^\prime = \mathcal{M}$ for an infinitesimal Lorentz transformation. Therefore, being invariant under infinitesimal Lorentz transformations, the amplitude (\ref{amplitude2bis}) will be invariant under finite Lorentz transformations.

\section{Unitarity and Cutkosky rules}\label{Section unitarity and Cutkosky rules}

In this section, we summarize briefly  the relation between unitarity and Cutkosky rules (see \cite{itzykson zuber,peskin,largest time equation} for a review). In particular, we will argue that, to complete the proof of the unitarity of nonlocal theories, we have to show that anomalous thresholds do not contribute to (\ref{Limit M}). This will be proved in Sec. \ref{section unitarity}.

The unitarity condition $S^\dagger S =1$ for	the $S$ matrix can be expressed in terms of the $T$ matrix as $T - T^\dagger  =  i	\, T^\dagger T$, where the $T$ matrix is defined by $S= 1 + i T$.
Taking the expectation value of the unitarity condition between an
initial incoming state $| a \rangle$ and an outgoing final state $
\langle b |$ for a given process $a \rightarrow b$, one has

\be T_{ba}  - T^*_{ab}=  i \,  \sum_c T^*_{c b} T_{c a}  \, \ee

Recasting the matrix elements of the $T$ matrix in terms of those
of the invariant scattering amplitude $\mathcal{M}$ as \mbox{
$T_{ab} = (2 \pi)^4  \, \mathcal{M}_{ab} \,	\delta^{(4)}\Big(\sum_i p_i - \sum_f p_f \Big)$}, where $p_i$ and $p_f$ are the initial an final external momenta, the unitarity
condition is finally expressed as

\be\label{unitarity condition M}
&& \mathcal{M}_{ba} - \mathcal{M}^*_{ab} = i \sum_c \,\, \mathcal{M}^*_{c
	b} \,\,\mathcal{M}_{c a} \,\, (2\pi)^4\,\, \delta^{(4)}(p_c-p_a) \, , \label{unitarityM}
\ee 
where $\mathcal{M}_{ba} = \langle b |
\mathcal{M} | a \rangle$ is the sum of all the connected amputated diagrams for the process $a \rightarrow b$ ( see \cite{peskin,itzykson zuber} for a review) and where we have neglected a global $\delta^{(4)}(\sum_i p_i - \sum_f p_f )$ multiplying both sides of Eq. (\ref{unitarity condition M}). The sum in $c$ is made on all possible real (nonvirtual) intermediate states, i.e., on those states that give nonzero amplitudes $\mathcal{M}_{bc} = \langle b |
\mathcal{M} | c \rangle$ and $\mathcal{M}_{ac} = \langle a |
\mathcal{M} | c \rangle$. The interpretation of (\ref{unitarity condition M}) is that at a Landau pole the amplitude $\mathcal{M}_{ba}$ has a branch cut singularity, which implies a discontinuity in its imaginary part above the corresponding threshold, which is related to the opening of a channel of production of intermediate real states $c$.

In order to prove the unitarity of the theory (\ref{phin2}), we must show that (\ref{unitarity condition M}) is satisfied for any process  $a \rightarrow b$. We have already shown that the  imaginary part of $\mathcal{M}$  at a specific Landau pole is given by the Cutkosky rules, replacing  each propagator that goes on shell with a term  $(- 2 \pi i) \,\delta(q^2-m^2)$ and  the integration contour $\mathcal{C}\times \mathbb{R}^3$ in the loops variables contained in $q$ with $\mathbb{R}^4$. The expression obtained in this way is usually referred as a cut diagram, since it is graphically represented by the same diagram as $\mathcal{M}$ in which the lines corresponding to the on shell propagators are cut  (see \cite{peskin,itzykson zuber,largest time equation} for review).

The Landau poles that are such that the corresponding cut diagram divide the original diagram in two are usually referred as normal thresholds. For such normal thresholds it is easy to show that (\ref{unitarity condition M}) is verified, since this is a straightforward consequence of the Cutkosky rules (see \cite{cutkosky} for review). Therefore, to complete our proof of the unitarity, we must show that only the Landau poles corresponding to normal thresholds contribute to the imaginary part  of $\mathcal{M}$. This is necessary since some Landau poles  are such that,  cutting the on shell propagators, the diagram is not divided in two, and the imaginary part of the amplitude cannot be recast as in (\ref{unitarity condition M}).  Such Landau poles are usually referred as anomalous thresholds.

Even in the case of a local theory, proving that anomalous thresholds do not contribute to (\ref{unitarity condition M})  is a difficult task, which requires methods based  on coordinate space analysis of diagrams, as the ‘largest time equation’ \cite{largest time equation}. However, the extension of such techniques to nonlocal theories is not obvious, because the largest time equation makes use of the fact that, in a local theory, the propagator can be divided in positive and negative energy parts, which is no longer the case in nonlocal theories. Yet, we can show that the diagrams that contribute to (\ref{unitarity condition M}) in the nonlocal case are all and only those that contribute in the local case; indeed there is no contribution from diagrams corresponding to anomalous thresholds. In fact,  we know that anomalous thresholds do not contribute in the local case, because the local theory is unitary. 

Before proceeding to this proof in Sec. \ref{section unitarity},  it is instructive to compute the limit (\ref{Limit M})  for some simple diagrams that do not have anomalous thresholds, proving their unitarity. This will be done in  Secs. \ref{Section one loop} and \ref{Section two loops}. There,  we will calculate $\mathcal{M}_{ba}$ from (\ref{amplitude2}) using the recursive equations (\ref{amplitude3}-\ref{amplitude3 integrated2}). Then we will obtain $\mathcal{M}^*_{ab}(E_h,\epsilon)$ by complex conjugation, and finally we will calculate $\mathcal{M}_{ab}(E_h,\epsilon)-\mathcal{M}^*_{ab}(E_h,\epsilon)$. Of course, the result will be in agreement with the Cutkosky rules.

\subsection{One-loop diagram}\label{Section one loop}
\begin{figure}
\begin{center}
\includegraphics[height=4cm]{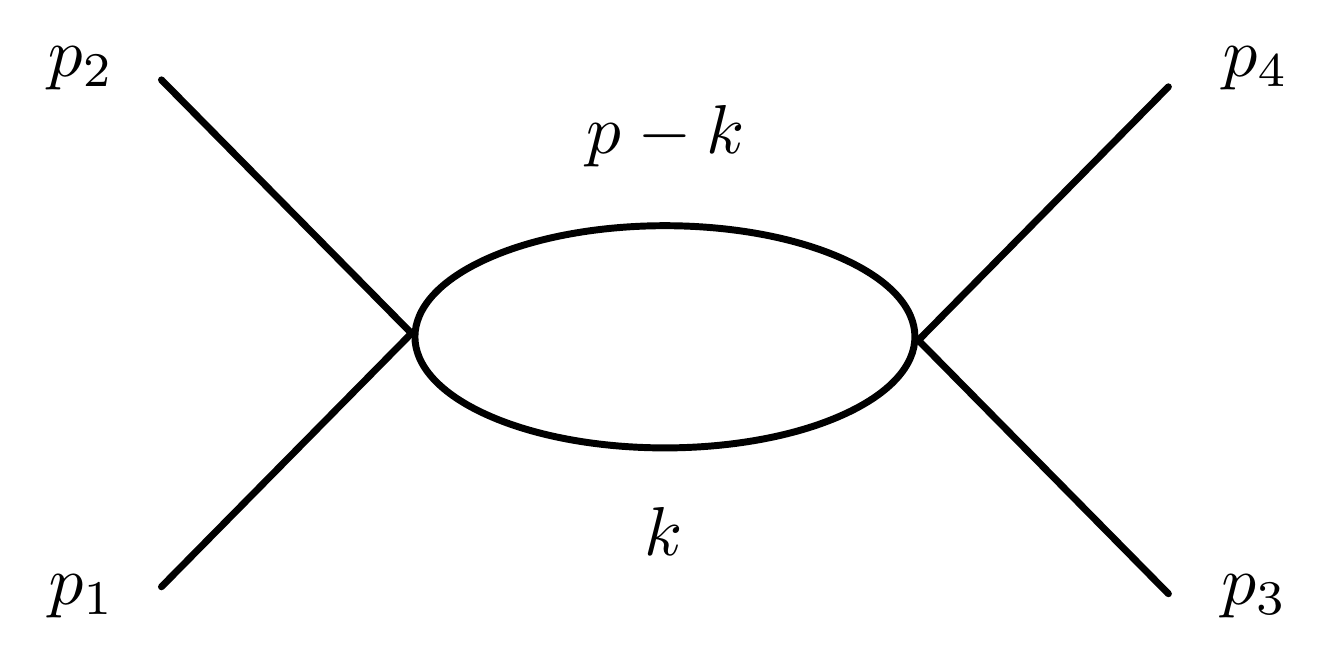}
\caption{Four-particle scattering amplitude in $\lambda \phi^4/4!$ theory at one-loop order.}
 \label{Fig2}
\end{center}
\end{figure}
Let us consider the one-loop four-particle scattering diagram in Fig.\ref{Fig2}, for a nonlocal scalar field with interaction $\lambda \phi^4/4!$. The corresponding
amputated amplitude in the Euclidean spacetime with purely
imaginary external energies is\footnote{Hereafter, we omit the indices $a b$ in $\mathcal{M}_{ab}$.}
\begin{equation}\label{amplitude 1loop 0}
\mathcal{M}(p_h,\epsilon) = - \frac{ \lambda^2}{2} \, \int_{(\mathcal{I}\times
\mathbb{R}^3)} \frac{i \, d^4 k}{(2\pi)^4}
\frac{\mathcal{V}(p_1,p_2,k,p-k)}{k^2 - m^2 + i \epsilon} \frac{
\mathcal{V}(p-k,k,p_3,p_4)}{(k-p)^2 - m^2 + i \epsilon} = - \frac{
\lambda^2}{2} \, \int_{(\mathcal{I}\times \mathbb{R}^3)} \frac{i
\, d^4 k}{(2\pi)^4} \frac{F_1(k,p_h)}{k^2 - m^2 + i \epsilon} \, ,
\end{equation}
where, in agreement with the notations
(\ref{amplitude3})-(\ref{definition FL1 5}),
\begin{equation}
F_1(k,p_h) \equiv F_1(k,p_1,p_2,p_3,p_4) = \frac{B(k,p_h)}{(k-p)^2
- m^2 + i \epsilon}  \, ,
\end{equation}
with $B(k,p_h) \equiv \mathcal{V}(p_1,p_2,k,p-k) \,
\mathcal{V}(p-k,k,p_3,p_4) $, and  $p = p_1+p_2=p_3+p_4$, where $p_1, p_2, p_3,p_4$ are the external momenta.

In order to continue analytically (\ref{amplitude 1loop 0}) to
real external energies, we should deform the integration contour $(\mathcal{I}\times
\mathbb{R}^3)$ around the poles of the propagators in (\ref{amplitude 1loop 0}) that pass through $\mathcal{I}$ when the energies $p^0_h$ become real, just as in (\ref{amplitude2bis}). Then, we can use the recursive procedure defined by  (\ref{amplitude3 integrated}) and (\ref{amplitude3 integrated2}), which make repeated use of the integration formula (\ref{formula integral 1}). Indeed, we must find the poles of the function
$F_1(k,p_h)$ in the variable $k^0$ and determine how they move in
the complex $k^0$ plane when the external energies are moved to
their physical values $p^0_1 \rightarrow E_1 \in
\mathbb{R}^+_0$, $p^0_2 \rightarrow E_2 \in \mathbb{R}^+_0$,
$p^0_3 \rightarrow E_3 \in \mathbb{R}^+_0$, $p^0_4 \rightarrow
E_4 \in \mathbb{R}^+_0$, so that $p^0 \rightarrow E \in \mathbb{R}^+_0$.

The poles of the first propagator are
\bea
\bar{k}^0_{1,2}
= \pm \sqrt{\vec{k}^2 + m^2 - i \epsilon}\, . 
\eea
Such poles do not depend on the
external energies and  remain always far from the imaginary
axis. Indeed, $\bar{k}^0_{1,2}$ do not pinch $\mathcal{I}$ and do not contribute to the sum of
residues in (\ref{formula integral 1}). However, such poles  overlap with those of the function $F_1(k,p_h)$  in the limit $i \epsilon \rightarrow 0$ for some values of the spatial loop momentum $\vec{k}$ and external momenta, constraining in between the
deformed contour $\mathcal{C}$, that will be then pinched.   That
implies the occurrence of a singularity at some threshold energy
and a corresponding branch cut discontinuity when $i \epsilon
\rightarrow 0$.

\begin{figure}
	\begin{center}
		\includegraphics[height=6cm]{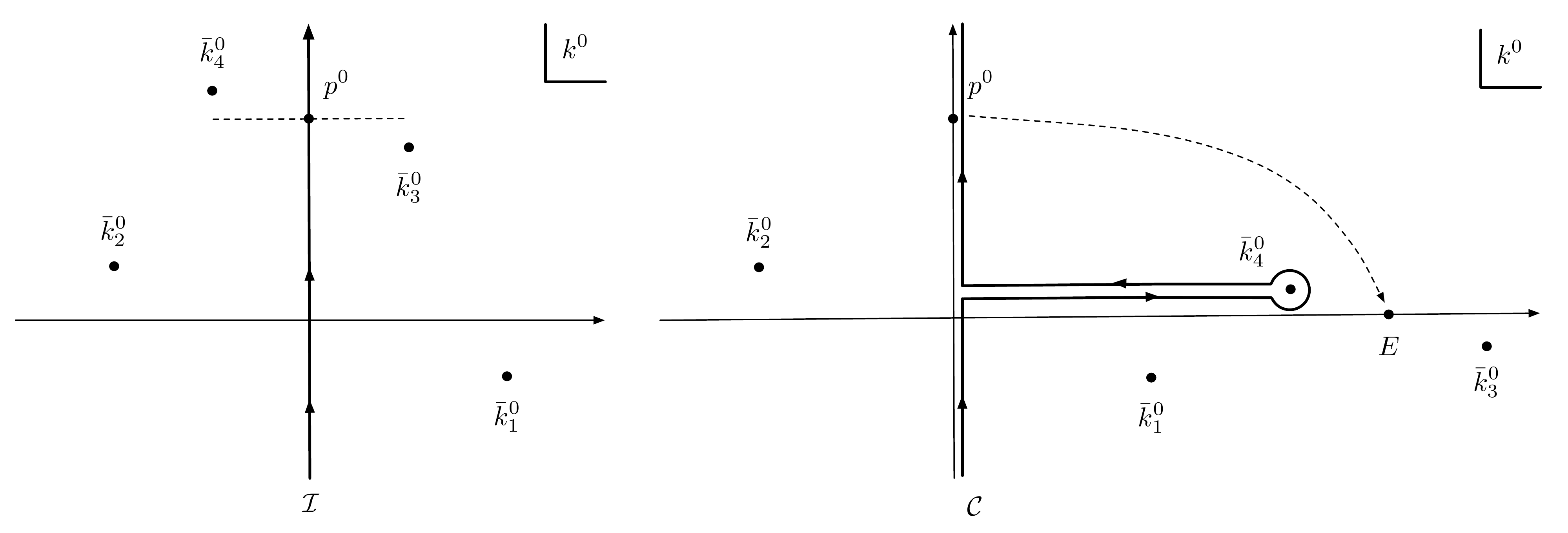}
		\caption{(Left) We plot the poles $\bar{k}^0_1, \ldots,\bar{k}^0_4$ on the complex $k^0$ plane,  when $p^0$ is purely imaginary. (Right) We plot the same poles for $p^0$ real and positive. Since $\bar{k}^0_1$ and $\bar{k}^0_2$ do not depend on $p^0$, their positions do not change when $p^0 \rightarrow E$. On the contrary, $\bar{k}^0_3$ and $\bar{k}^0_4$ move to the right, and $\bar{k}^0_4$ passes through the imaginary axis $\mathcal{I}$ for some values of $\vec{k}$. We also plot the contour $\mathcal{C}$, which is obtained deforming $\mathcal{I}$ around $\bar{k}^0_4$.}
		\label{Fig3}
	\end{center}
\end{figure}

The poles of $F_1(k,p_h)$ are given by the two zeros of its
denominator, which are 
\bea
\bar{k}^0_{3,4} = p^0 \pm
\sqrt{(\vec{k}-\vec{p})^2+m^2-i\epsilon }. 
\eea
These two poles are at
the right and  left of $p^0$, which is initially purely
imaginary; indeed, $\bar{k}^0_{3}$ is at the right and
$\bar{k}^0_{4}$ is at the left of the imaginary axis of the $k^0$
plane, for purely imaginary $p^0$. When $p^0$ is moved to its
physical real and positive value $E$, the two poles
$\bar{k}^0_{3,4}$ move to the right and it happens that the pole
$\bar{k}^0_{4}$ passes through the imaginary axis for some values
of the loop momenta $\vec{k}$  and the external energy (see Fig.\ref{Fig3}).
Indeed, the integration contour $\mathcal{C}$ is obtained
deforming the imaginary axis $\mathcal{I}$ around the pole
$\bar{k}^0_{4}$ when such pole passes through
$\mathcal{I}$, i.e.,  when $\Re\left\{\bar{k}^0_{4}\right\} > 0$, where $\Re\left\{a\right\}$ is the real part of the complex number $a\in \mathbb{C}$.

Having defined the deformed integration contour, we can write the analytical  continuation of the amplitude (\ref{amplitude 1loop 0}) as

\begin{equation}\label{amplitude 1loop 1}
\mathcal{M}(p_h,\epsilon) = - \frac{ \lambda^2}{2} \, \int_{(\mathcal{C}\times
	\mathbb{R}^3)} \frac{i \, d^4 k}{(2\pi)^4} \frac{1}{k^2 -
	m^2 + i \epsilon} \frac{B(k,p_h)}{(k-p)^2
	- m^2 + i \epsilon}  = - \frac{ \lambda^2}{2} \, \int_{(\mathcal{C}\times
\mathbb{R}^3)} \frac{i \, d^4 k}{(2\pi)^4} \frac{F_1(k,p_h)}{k^2 -
m^2 + i \epsilon} \, ,
\end{equation}
where now $p^0 = p^0_1+p^0_2= E \in \mathbb{R}^+_0$ is real and positive.

According to (\ref{formula integral 1}), we can divide this
integral in two parts, separating the contribution of the
residuals from the integral on $\mathcal{I}$, which gives

\begin{equation}\label{amplitude 1loop 3}
\mathcal{M}(p_h,\epsilon)  = - \frac{ \lambda^2}{2}    \left[
\int_{(\mathcal{I}\times \mathbb{R}^3)} \frac{i \, d^4
k}{(2\pi)^4} \frac{F_1(k,p_h)}{k^2 - m^2 + i \epsilon} +
\int_{(\mathbb{R}^3)} \frac{i \, d^3 k}{(2\pi)^4} \, 2\pi i \,
\sum_l \, {\rm Res} \left\{\frac{F_1(k,p_h)}{k^2 - m^2 + i
\epsilon},\bar{k}^0_{l}\right\} \right] ,
\end{equation}
where the sum is extended to all the residues $\bar{k}^0_{l}$ that
pass through $\mathcal{I}$ when $p^0$ passes from purely imaginary
to real and positive values. Therefore, the only pole that
contributes to the sum of the residues in (\ref{amplitude 1loop 3})
is \mbox{$\bar{k}^0_4 = p^0 - \sqrt{(\vec{k}-\vec{p})^2+m^2-i\epsilon
}$}, which passes from the left to the right of $\mathcal{I}$ when
$p^0 \rightarrow E$, for values of the external energies and
internal loop momenta $\vec{k}$ such that
$\Re\left\{\bar{k}^0_4\right\}
> 0$. Thus, Eq.(\ref{amplitude 1loop 3}) gives

\bea
&&
\hspace{-0.5cm}
\mathcal{M}(p_h,\epsilon)  = - \frac{ \lambda^2}{2}    \left[
\int_{(\mathcal{I}\times \mathbb{R}^3)} \frac{i \, d^4 k}{(2\pi)^4} \frac{F_1(k)}{k^2 - m^2 + i \epsilon} +
\int_{(\mathbb{R}^3)} \frac{i \, d^3 k}{(2\pi)^4}  \left(-2\pi i\right) \!
\left(\frac{\sigma(\Re\left\{k^0\right\})}{2\sqrt{(\vec{k}-\vec{p})^2+m^2-i\epsilon
}} \,\frac{B(k,p_h)}{k^2 - m^2 + i
\epsilon}\right)\Bigg|_{k^0=\bar{k}^0_4} \right] , \nonumber \\
&& \label{amplitude1loop4}
\eea
where
\begin{equation}
\sigma(x) = \left\{ \begin{array}{ll} 1, \quad x>0\\
1/2, \quad x=0\\
0, \quad x<0
\end{array}\right. \,.
\end{equation}
The $\sigma(\Re(\bar{k}^0_4))$ in (\ref{formula integral 1})
implies that when the pole $\bar{k}^0_4$ is in the first or fourth
quadrant of the complex plane it contributes a $2 \pi i $  times
the residue. When it is on the imaginary axis $\mathcal{I}$, it
contributes just the half, while when it is in the second and
third quadrant it does not contribute. Moreover, when the pole is
on the imaginary axis, the integral over $\mathcal{I}$ is intended
as the principal part. Therefore, the support of
$\sigma(\Re(\bar{k}^0_4))$ defines the region of integration in
$d^3k$. Furthermore, 
$\sigma(\Re(\bar{k}^0_4))=1/2$ on a
subset of $\mathbb{R}^3$ with zero measure; thus, such subset does
not contribute to the second integral in (\ref{formula integral 0}).

In what follows, we will make use of the following relation:
\be \label{delta complessa}
\int_{\mathbb{R}^3}  d^3 k \frac{f(k,h)}{2 \sqrt{(\vec{k}-\vec{h})^2+m^2}} \, \Bigg|_{k^0 \equiv h^0 \mp \sqrt{(\vec{k}-\vec{h})^2+m^2}} =  \int_{\mathbb{R}^4} d^4 k \, \sigma\left(\pm \left(h^0-k^0\right)\right) \, \delta((h-k)^2-m^2) f(k,h) \, ,
\ee
where the $\sigma\left(\pm \left(h^0-k^0\right)\right)$ function selects the correct root $h^0 \mp \sqrt{(\vec{k}-\vec{h})^2+m^2}$ of the delta function. Let us consider the last integral in
(\ref{amplitude1loop4}). Since  $(\vec{k}-\vec{p})^2+m^2 \neq 0$, we can neglect the $i \epsilon$ term in $\sqrt{(\vec{k}-\vec{p})^2+m^2-i\epsilon }$, since we are interested in the limit $\epsilon \rightarrow 0$ of the amplitude. Therefore, according to (\ref{delta complessa}), in the limit of small $\epsilon$ one has

\be
\label{amplitude 1loop 5}
\int_{(\mathbb{R}^3)} \!\!\!\!   d^3k  \left(\frac{\sigma(\Re\left\{k^0\right\})}{2\sqrt{(\vec{k}-\vec{p})^2+m^2-i\epsilon
}} \,\,\frac{B(k,p_h)}{k^2 - m^2 + i
	\epsilon}\right) \! \Bigg|_{k^0=\bar{k}^0_4}  \!\!\! = 
 \int_{(\mathbb{R}^4)} \!\!\!
d^4k  \, \frac{B(k,p_h)}{k^2 - m^2 + i \epsilon} \, 
\sigma(k^0) \sigma(p^0-k^0)\,\delta((p-k)^2-m^2)   ,  
\ee
where we have maintained the term $i \epsilon$ in the propagator $1/(k^2 - m^2 + i \epsilon)$ since this propagator can go on shell.
Therefore, the final expression of the analytic continuation (\ref{amplitude 1loop 1}) is obtained by replacing (\ref{amplitude 1loop 5}) in (\ref{amplitude1loop4}), obtaining
\be
 &&\hspace{-4cm}\mathcal{M}(p_h,\epsilon)  = - \frac{ \lambda^2}{2}    
 \left[
\int_{(\mathcal{I}\times \mathbb{R}^3)} \frac{i \, d^4
k}{(2\pi)^4} \, B(k,p_h)\, \frac{1}{k^2 - m^2 + i \epsilon}
\frac{1}{(k-p)^2 - m^2 + i \epsilon} +  \right. \nonumber\\ 
&& \hspace{-1.6cm}
\left. +  \int_{(\mathbb{R}^4)} \frac{i \, d^4 k}{(2\pi)^4}  \,
\,\frac{B(k,p_h) }{k^2 - m^2 + i \epsilon} \, (-2\pi i)
\, \sigma(k^0) \sigma(p^0-k^0) \,\delta((p-k)^2-m^2) \right] \, .
\label{amplitude 1loop 6}
\ee
We note that the first integral in (\ref{amplitude 1loop 6}) is of the same kind of the integral as (\ref{Mba}); therefore, following the steps (\ref{Mba})-(\ref{Mab conj3}), one easily realizes that such integral  is real in the limit $i \epsilon \rightarrow 0$, and it does not contribute to $\mathcal{M} -\mathcal{M}^*$. Notice that  $k^2 - m^2 \neq 0$ and $(k-p)^2 - m^2  \neq 0$ in the first integral of (\ref{amplitude 1loop 6}) because $k \in \mathcal{I}\times \mathbb{R}^3$ and  $p \in
	\mathbb{R}^+_0 \times \mathbb{R}^3$, so that the integrand is not singular on the integration contour.

We can now evaluate $\mathcal{M}(p_h,\epsilon) -\mathcal{M}(p_h,\epsilon) ^*$ in the limit $i \epsilon \rightarrow 0$ and verify the validity of the Cutkosky rules for our nonlocal theory. We have
\bea\label{unitarity 1loop 1a}
&& \hspace{-0.7cm}
\mathcal{M}(p_h,\epsilon)  -\mathcal{M}(p_h,\epsilon) ^* = \nonumber \\
&& = - \frac{ \lambda^2}{2}  \label{unitarity 1loop 1b} \int_{(\mathbb{R}^4)} \frac{i \, d^4 k}{(2\pi)^4} \,
B(k,p_h) \, (-2\pi i)\,\sigma(k^0) \sigma(p^0-k^0)\,\delta((p-k)^2-m^2) \,
\left[\frac{ 1}{k^2 - m^2 + i \epsilon}-\frac{ 1}{k^2 - m^2 - i
\epsilon}\right] .
\eea
In order to recast
the second integral of (\ref{unitarity 1loop 1b}), we just note
that it is performed on the real $\mathbb{R}^4$ space; thus, one is
allowed to use the formula
\begin{equation}\label{delta}
\lim_{\epsilon \rightarrow 0}\left\{\frac{ 1}{k^2 - m^2 + i
\epsilon}-\frac{ 1}{k^2 - m^2 - i \epsilon}\right\} = (-2\pi i) \,
\delta(k^2 - m^2) \, ,
\end{equation}
obtaining

\bea
\label{unitarity1loop2}\hspace{-1cm}
&&\lim_{\epsilon \rightarrow 0} \left\{\mathcal{M}(p_h,\epsilon)  -\mathcal{M}(p_h,\epsilon) ^* \right\}= - \frac{ \lambda^2}{2}
\int_{(\mathbb{R}^4)} \frac{i \, d^4 k}{(2\pi)^4} \, B(k,p_h) \,
(-2\pi i)^2\,  \sigma(p^0-k^0) \,\delta((p-k)^2-m^2) \,  \sigma(k^0)
\delta(k^2 - m^2) \,,
\eea
where we have used the fact that $k^0 \in \mathbb{R}$ when $i \epsilon \rightarrow 0$, which implies that $\sigma(\Re\left\{k^0\right\})=\sigma(k^0)$.  The $\sigma(k^0)$ in (\ref{unitarity1loop2}) selects the positive $k^0$ root  of the delta functions, corresponding to the physical energy of real intermediate states when the loop momenta are on shell.

Equation (\ref{unitarity1loop2}) is just the nonlocal version of the standard local Cutkosky rules, which state that the discontinuity of the amplitude at some cut is given by replacing each propagator
\bea
\frac{i}{p^2-m^2+i\epsilon} \qquad \longrightarrow \qquad  (- 2 \pi i)\, \sigma(p^0)\,
\delta(p^2-m^2+i\epsilon) \, . 
\label{cutR}
\eea
The only difference is that, in the
case of a nonlocal theory, one has also to replace the integration
contour $\mathcal{C}\times \mathbb{R}^3$ with $\mathbb{R}^4$.

Finally, we note that, in all the theories we are interested in, the vertex function is such that $\mathcal{V}=1 $  when the corresponding momenta are on shell. For instance, that happens when the function $H(z)$ in  (\ref{vertex V definition}) is such that \mbox{$H(p^2= m^2)= 0$}. In such theories, one has  $B(k,p_h)=1$ when all the momenta are on shell, and (\ref{unitarity 1loop 4}) simplifies  
\bea
\label{unitarity 1loop 4}
\mathcal{M}(p_h,\epsilon)  -\mathcal{M}(p_h,\epsilon) ^* = - \frac{ \lambda^2}{2}
\int_{(\mathbb{R}^4)} \frac{i \, d^4 k}{(2\pi)^4}  \,
(-2\pi i)\, \sigma(k^0) \,\delta((k-p)^2-m^2) \, (-2\pi i)\, \sigma(p^0)\,
\delta(k^2 - m^2) \, .
\eea
%
For these special but very common \cite{Krasnikov, kuzmin, modesto, modestoLeslaw, Modesto:2015lna, piva} theories we can replace (\ref{cutR}) with
\bea
\frac{i \, e^{- H(p^2)}}{p^2-m^2+i\epsilon} \qquad \longrightarrow \qquad  (- 2 \pi i) \, \sigma(p^0)\,
\delta(p^2-m^2+i\epsilon) \, . 
\eea
%

\subsection{Two-loop diagram}\label{Section two loops}

\begin{figure}
\begin{center}
\includegraphics[height=4cm]{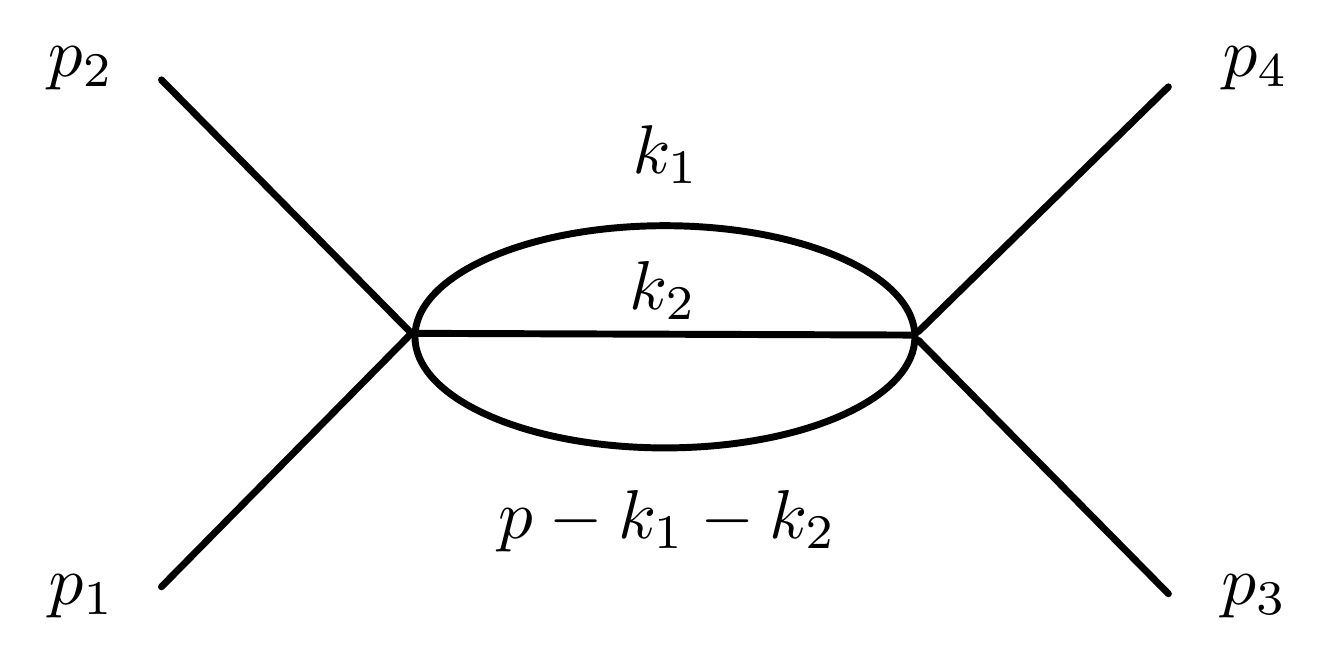}
\caption{Four-particle scattering amplitude at two loops in $\lambda \phi^5$ theory.}
 \label{Fig5}
\end{center}
\end{figure}

In this section, we study the two-loop diagram in Fig.\ref{Fig5}  for an Euclidean scalar field theory with  $\lambda \phi^5/5!$ interaction. The  
complex amplitude for this process  is

\be
&& \label{amplitude2loop0a}
\mathcal{M}(p_h,\epsilon) = - \frac{ \lambda^2}{2} \, \int_{(\mathcal{I}\times
\mathbb{R}^3)^2} \, \frac{i \, d^4 k_1}{(2\pi)^4}\,\frac{i \, d^4
k_2}{(2\pi)^4}\, \frac{1}{k_1^2 - m^2 + i \epsilon}\,
\frac{\mathcal{V}(p,k_1,k_2,p-k_1-k_2)}{k_2^2 - m^2 + i \epsilon}\,
\frac{\mathcal{V}(k_1,k_2,p-k_1-k_2,p)}{(k_1+k_2-p)^2 - m^2 + i \epsilon} \nonumber \\
&&\label{amplitude2loop0b}
\hspace{1.45cm} 
= - \frac{ \lambda^2}{2} \, \int_{(\mathcal{I}\times
\mathbb{R}^3)} \frac{i \, d^4 k_2}{(2\pi)^4}  \frac{\tilde{F}_2(k_2,p_h)
}{k_2^2 - m^2 + i \epsilon} \, ,
\ee
where we are assuming purely imaginary
external energies $p^0_1, \, p^0_2, \, p^0_3, \, p^0_4$, and
where we introduced the definitions
\be
&& \label{amplitude 2loop 1}
\tilde{F}_2(k_2,p_h) = \int_{(\mathcal{I}\times \mathbb{R}^3)} \, \frac{i
\, d^4 k_1}{(2\pi)^4}\, \frac{1}{k_1^2 - m^2 + i \epsilon}\,
\frac{\tilde{B}(k_1, k_2, p_h)}{(k_1+k_2-p)^2 - m^2 + i \epsilon} = 
\int_{(\mathcal{I}\times \mathbb{R}^3)} \, \frac{i \, d^4
k_1}{(2\pi)^4}\, \frac{\tilde{F}_1(k_2,k_1,p_h)}{k_1^2 - m^2 + i \epsilon}
\,, \nonumber \\
&& \label{amplitude 2loop 2}
\tilde{F}_1(k_2,k_1,p_h) =  \frac{\tilde{B}(k_1,k_2,p_h)}{(k_1+k_2-p)^2 -
m^2 + i \epsilon}  \,,
\ee
with  
\be
\tilde{B}(k_1,k_2,p_h) \equiv \mathcal{V}(p,k_1,k_2,p-k_1-k_2)   \,
\mathcal{V}(k_1,k_2,p-k_1-k_2,p) \, ,
\ee 
and $p= p_1+p_2 = p_3+p_4$ as the external momentum.

The expressions (\ref{amplitude2loop0b}) and (\ref{amplitude 2loop 2}) are well defined when the external energies are purely
imaginary because in that case the poles of the propagators are far
from the integration contour $(\mathcal{I}\times \mathbb{R}^3)^2$.
Therefore, to find the physical amplitude and prove the Cutkosky
rules, we have to continue {\em  analytically} (\ref{amplitude2loop0b})
and (\ref{amplitude 2loop 2}) to real and positive external energies. This is done deforming
the integration contour around the poles that pass through
$(\mathcal{I}\times \mathbb{R}^3)^2$ when  $p^0_1 \rightarrow E_1 \in
\mathbb{R}^+_0$, $p^0_2 \rightarrow E_2 \in \mathbb{R}^+_0$,
$p^0_3 \rightarrow E_3 \in \mathbb{R}^+_0$, and $p^0_4 \rightarrow
E_4 \in \mathbb{R}^+_0$. 
The physical amplitude becomes
\begin{equation}\label{amplitude 2loop 3}
\mathcal{M}(p_h,\epsilon) = -  \frac{ \lambda^2}{2} \,
\int_{(\mathcal{C}_2\times \mathbb{R}^3)} \frac{i \, d^4
k_2}{(2\pi)^4}  \frac{\tilde{F}_2(k_2,p_h) }{k_2^2 - m^2 + i \epsilon} \, ,
\end{equation}
with
\begin{equation}\label{amplitude 2loop 3b} \tilde{F}_2(k_2,p_h) 
=\int_{(\mathcal{C}_1\times \mathbb{R}^3)} \, \frac{i \, d^4	k_1}{(2\pi)^4}\, \frac{1}{k_1^2 - m^2 + i \epsilon} \frac{\tilde{B}(k_1,k_2,p)}{(k_1-(p-k_2))^2 -	m^2 + i \epsilon}=
\int_{(\mathcal{C}_1\times \mathbb{R}^3)} \, \frac{i \, d^4	k_1}{(2\pi)^4}\, \frac{\tilde{F}_1(k_2,k_1,p_h)}{k_1^2 - m^2 + i \epsilon}\,,
\end{equation}
where the contours $(\mathcal{C}_1\times \mathbb{R}^3)$ and
$(\mathcal{C}_2\times \mathbb{R}^3)$ are determined by the
analysis of the poles of $\tilde{F}_1(k_2,k_1,p_h)$ and $\tilde{F}_2(k_2,p_h)$ as follows.

\begin{figure}
	\begin{center}
		\includegraphics[height=6cm]{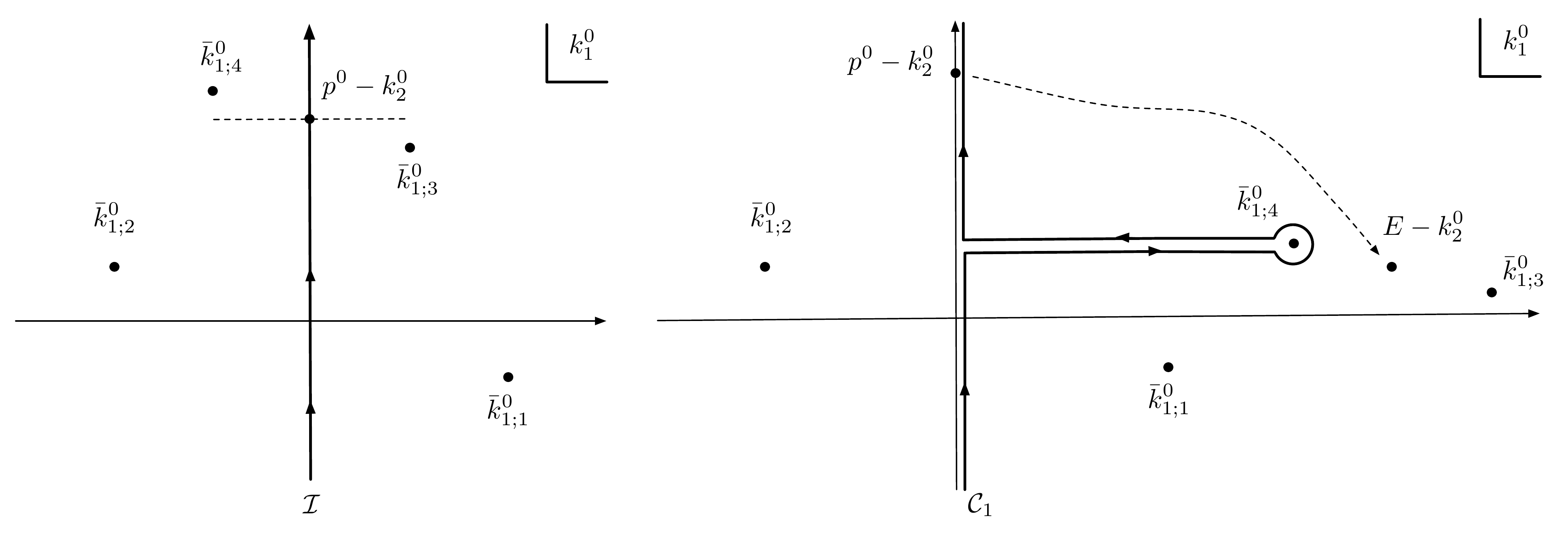}
		\caption{(Left) We plot the poles $\bar{k}^0_{1;1}, \ldots,\bar{k}^0_{1;4}$ on the complex $k^0_1$ plane,  when $p^0$ is purely imaginary, which implies that $k^0_2$ is purely imaginary too. (Right) We plot the same poles for $p^0$ real and positive. Since $\bar{k}^0_{1;1}$ and $\bar{k}^0_{1;2}$ do not depend on $p^0$, their positions do not change when $p^0 \rightarrow E$. On the contrary, $\bar{k}^0_{1;3}$ and $\bar{k}^0_{1;4}$ move to the right, and $\bar{k}^0_4$ passes through the imaginary axis $\mathcal{I}$ for some values of $\vec{k}_1$. We also plot the contour $\mathcal{C}_1$, which is obtained deforming $\mathcal{I}$ around $\bar{k}^0_{1;4}$.}
		\label{Fig6}
	\end{center}
\end{figure}

Let us start from the expression of $\tilde{F}_2(k_2,p_h)$ in (\ref{amplitude 2loop 3b}). In order to apply the integral formula (\ref{formula integral 1}), we have to find the integration contour $(\mathcal{C}_1\times \mathbb{R}^3)$ by determining which one of the poles  of $\tilde{F}_1(k_2,k_1,p_h)$ moves through the imaginary axis of the $k^0_1$ plane when $p^0 \rightarrow E \in \mathbb{R}^+_0$. To proceed, we can use the analogy with the one-loop case as a guideline.
In fact, (\ref{amplitude 2loop 3b}) is the same as
(\ref{amplitude 1loop 1}) with the replacements 
\be 
k \rightarrow k_1 \,,  \quad 
p \rightarrow p-k_2 \, , \quad 
F_1(k,p) \rightarrow \tilde{F}_1(k_1,p-k_2) \,  , \quad 
 B(k,p_h) \rightarrow \tilde{B}(k_1,k_2,p_h). 
\ee
The only difference concerns the positions of the poles in the $k^0_1$ complex plane, which now also depend on $k_2$. 
The roots of $k_1^2 - m^2 + i \epsilon = 0$, i.e., 
\be
\bar{k}^0_{1;1,2}= \pm \sqrt{\vec{k}_1^2+m^2-i \epsilon}
\ee
 do not depend on $p^0$ and they are always far from the imaginary axes. The poles of  $\tilde{F}_2(k_2,p_h)$ are 
 \be
\bar{k}^0_{1;3,4}= p^0-k^0_2 \pm \sqrt{(\vec{k}_1+\vec{k}_2-\vec{p})^2+m^2-i \epsilon} \, , 
\ee
and they are at the left and right of the imaginary axes when $p^0$ is purely imaginary, because in that case also $k^0_2$ is purely imaginary, since we integrate it on $\mathcal{I}$ [see (\ref{amplitude2loop0a})]. Both the poles $\bar{k}^0_{1;3,4}$ move to the right when $p^0 \rightarrow E \in \mathbb{R}^+_0$,  so that the only pole that might cross the imaginary axes is 
 \be
 \bar{k}^0_{1;4}= p^0-k^0_2 - \sqrt{(\vec{p}-\vec{k}_1-\vec{k}_2)^2+m^2-i \epsilon} \, , 
 \ee
 and this happens only when $\Re\left\{\bar{k}^0_{1;4}\right\}>0$, see Fig.\ref{Fig6}. Therefore, $(\mathcal{C}_1\times \mathbb{R}^3)$ is obtained deforming $(\mathcal{I}\times \mathbb{R}^3)$ around $\bar{k}^0_{1;4}$ when it crosses the imaginary axes. Now using (\ref{amplitude1loop4}) we can directly derive  the expression of $\tilde{F}_2(k_2, p_h)$ as

\be\label{ftilde}
&& \tilde{F}_2(k_2,p_h) =  \int_{(\mathcal{I}\times \mathbb{R}^3)} \frac{i \,
	d^4 k_1}{(2\pi)^4}  \, \frac{1}{k_1^2 - m^2
	+ i \epsilon} \frac{ \tilde{B}(k_1,k_2,p)}{(k_1+k_2-p)^2 - m^2 + i \epsilon} 
\nonumber \\
&& \hspace{1.6cm}
+ \int_{(\mathbb{R}^3)} \frac{i \, d^3 k_1}{(2\pi)^4} \, \left(-2\pi i\right) \,
\left(\frac{\sigma(\Re\left\{k^0_1\right\})}{2\sqrt{(\vec{p}-\vec{k}_1-\vec{k}_2)^2+m^2-i\epsilon
}} \,\,\frac{B(k,p_h)}{k_1^2 - m^2 + i	\epsilon}\right)\Bigg|_{k^0_1=\bar{k}^0_{1;4}}  \, ,
\ee

Therefore, (\ref{amplitude 2loop 3}) becomes
\be
&& \hspace{-1cm}
\mathcal{M}(p_h,\epsilon) =  - \frac{ \lambda^2}{2} \, \int_{(\mathcal{C}_2\times
\mathbb{R}^3)} \frac{i \, d^4 k_2}{(2\pi)^4}  \frac{1 }{k_2^2 -
m^2 + i \epsilon} \left\{ \int_{(\mathcal{I}\times \mathbb{R}^3)} \frac{i
\, d^4 k_1}{(2\pi)^4}  \, \frac{1}{k_1^2 -
m^2 + i \epsilon} \frac{\tilde{B}(k_1,k_2,p)}{(k_1+k_2-p)^2 - m^2 + i \epsilon}  \right.
\nonumber \\
&& 
\label{amplitude 2loop 5}
\hspace{1.3cm}
+\left. \int_{(\mathbb{R}^3)} \frac{i \, d^3 k_1}{(2\pi)^4} \, \left(-2\pi i\right) \,
\left(\frac{\sigma(\Re\left\{k^0_1\right\})}{2\sqrt{(\vec{p}-\vec{k}_1-\vec{k}_2)^2+m^2-i\epsilon
}} \,\,\frac{B(k,p_h)}{k_1^2 - m^2 + i	\epsilon}\right)\Bigg|_{k^0_1=\bar{k}^0_{1;4}} \right\}\, .
\ee
We stress that the delta function in (\ref{amplitude 2loop 5}) has to be interpreted according  to 
(\ref{delta complessa}) when $k^0_2 \in I$ is purely imaginary.
Now we have to analyze the poles of the integrand in (\ref{amplitude 2loop 5}) in order to find  contour $(\mathcal{C}_2\times \mathbb{R}^3)$.   For the first integral in (\ref{amplitude 2loop 5}), we have

\be
&&\hspace{-1cm} 
 \int_{(\mathcal{C}_2\times \mathbb{R}^3)} \frac{i \, d^4
k_2}{(2\pi)^4}  \frac{1 }{k_2^2 - m^2 + i \epsilon}
\int_{(\mathcal{I}\times \mathbb{R}^3)} \frac{i \, d^4
k_1}{(2\pi)^4} \,  \frac{1}{k_1^2 - m^2 + i
\epsilon} \frac{\tilde{B}(k_1,k_2,p)}{(k_1+k_2-p)^2 - m^2 + i \epsilon}  \nonumber \\
&& \hspace{-1cm} 
=\int_{(\mathcal{I}\times \mathbb{R}^3)} \frac{i \, d^4
k_1}{(2\pi)^4}\frac{1}{k_1^2 - m^2 + i \epsilon} \,
\int_{(\mathcal{C}_2\times \mathbb{R}^3)} \frac{i \, d^4
k_2}{(2\pi)^4}\,  \frac{1 }{k_2^2 - m^2 + i
\epsilon}\, \frac{\tilde{B}(k_1,k_2,p)}{(k_1+k_2-p)^2 - m^2 + i \epsilon}  \nonumber \\
&& \hspace{-1cm} 
\label{amplitude 2loop 6}
=\int_{(\mathcal{I}\times \mathbb{R}^3)} \frac{i \, d^4
k_1}{(2\pi)^4}\frac{1}{k_1^2 - m^2 + i \epsilon}
\left[\int_{(\mathcal{I}\times \mathbb{R}^3)} \frac{i \, d^4
k_2}{(2\pi)^4} \,  \frac{1}{k_2^2 - m^2 + i
\epsilon} \frac{\tilde{B}(k_1,k_2,p)}{(k_1+k_2-p)^2 - m^2 + i \epsilon} \right.
\nonumber \\
&&  
\label{amplitude 2loop 7}
\hspace{3.5cm}
\left.+  
\int_{(\mathbb{R}^3)} \frac{i \, d^3 k_2}{(2\pi)^4} \, \left(-2\pi i\right) \,
\left(\frac{\sigma(\Re\left\{k^0_2\right\})}{2\sqrt{(\vec{p}-\vec{k}_1-\vec{k}_2)^2+m^2-i\epsilon
}} \,\,\frac{B(k,p_h)}{k_2^2 - m^2 + i	\epsilon}\right)\Bigg|_{k^0_2=\bar{k}^0_{2;4}}
\right] \,,
\ee
where in the first equality we have just inverted the order of integration, while the integral in $d^4k_2$ has been obtained using (\ref{ftilde}) with the replacement $k_1\leftrightarrow k_2$, with the definition
\be
\bar{k}^0_{2;4}= p^0-k^0_1 - \sqrt{(\vec{p}-\vec{k}_1-\vec{k}_2)^2+m^2-i \epsilon} \, . 
\ee
We note that the two integrals  in (\ref{amplitude 2loop 7}) are real in the limit $\epsilon \rightarrow 0$, therefore, they will not contribute to $\mathcal{M}-\mathcal{M}^*$. In fact, for what concerns the first integral in (\ref{amplitude 2loop 7}), one can easily repeat the steps (\ref{Mba})-(\ref{Mab conj3}) and conclude that  (\ref{amplitude 2loop 7}) is real for $\epsilon \rightarrow 0$. For the second integral in (\ref{amplitude 2loop 7}), we note that  $k_1 \in \mathcal{I}\times \mathbb{R}^3$, $k^0_2=\bar{k}^0_{2;4}$ and $\vec{k}_2 \in  \mathbb{R}^3$, so that $k_1^2 - m^2  \in \mathbb{R}$ while $k_2^2 - m^2 \in \mathbb{R}$ when $\epsilon \rightarrow 0$, and therefore such integral is real for $\epsilon \rightarrow 0$.

Let us now consider the second integral in (\ref{amplitude 2loop 5}). We have to evaluate this integral according to the integration formula (\ref{formula integral 1}), that is,

\begin{equation}\label{integral 2loop 2 1}
\int_{(\mathcal{C}_2\times \mathbb{R}^3)} \frac{i \, d^4
	k_2}{(2\pi)^4}  \frac{ G(k_2,p_h) }{k_2^2 - m^2 + i \epsilon} =
\int_{(\mathcal{I}\times \mathbb{R}^3)} \frac{i \, d^4
	k_2}{(2\pi)^4}  \frac{ G(k_2,p_h) }{k_2^2 - m^2 + i \epsilon}+ 
\int_{(\mathbb{R})^3}\frac{i \, d^3 k_2}{(2\pi)^4} \, (2 \pi i ) \, \sum_h
{\rm Res} \left\{\frac{ G(k_2,p_h) }{k_2^2 - m^2 + i \epsilon},\bar{k}^0_{2;h}\right\} \, , 
\end{equation}
where  we have defined

\be
&& \hspace{-1cm}
G(k_2,p_h)=\int_{(\mathbb{R}^3)} \,\frac{i \, d^3 k_1}{(2\pi)^4}  \! \left(
\frac{ (-2\pi i)\sigma\left(\Re\left\{k^0_1\right\}\right) }{2\sqrt{(\vec{k}_1+\vec{k}_2-\vec{p})^2+m^2-i\epsilon}}\,\,
\frac{\,\tilde{B}(k_1,k_2,p) }{k_1^2 - m^2 + i	\epsilon}\right)\Bigg|_{ k_1^0=\bar{k}^0_{1;4}} \nonumber
\\
&& \hspace{0.5cm}
\label{integral II 0}
=\int_{(\mathbb{R}^3)} \,\frac{i \, d^3 k_1}{(2\pi)^4} \! 
\left(\frac{(-2\pi i)}{2\sqrt{(\vec{k}_1+\vec{k}_2-\vec{p})^2+m^2-i\epsilon}}\,\,
\frac{\sigma\left(\Re\left\{k^0_1\right\}\right) }{k_1^0+\sqrt{\vec{k_1}^2+m^2   - i\epsilon}} \, \frac{\,\tilde{B}(k_1,k_2,p) }{k_1^0-\sqrt{\vec{k_1}^2+m^2 - i \epsilon}}\right)\Bigg|_{k_1^0=\bar{k}^0_{1;4}}\, .
\ee
The sum on the rhs of (\ref{integral 2loop 2 1}) is limited to the residues corresponding to the poles $\bar{k}^0_{2;h}$ that pass through the imaginary axis of the $k^0_2$ plane when we take real positive external energies, i.e., $p^0 \rightarrow E \in \mathbb{R}^+_0$, and  the contour $(\mathcal{C}_2\times \mathbb{R}^3)$ in (\ref{integral 2loop 2 1}) is obtained deforming $(\mathcal{I}\times \mathbb{R}^3)$ around the same poles. 

Therefore, to compute (\ref{integral 2loop 2 1}) we have to find the poles of the function $G(k_2,p_h))/(k_2^2-m^2+i\epsilon)$ and study their dependence on $p^0$. The first two poles are given by  the zeros of the denominator and they are located in
\be
\bar{k}_{2;1,2}^0 = \pm \sqrt{\vec{k}^2_2 + m^2-i\epsilon}. 
\ee
Such poles are in the second and fourth quadrant of the $k_2$ complex plane, far away from the imaginary axes $\mathcal{I}$. Moreover, $\bar{k}_{2;1,2}^0$ never cross $\mathcal{I}$ because they do not depend on $p^0$. Therefore, they do not contribute to the sum of residues in (\ref{integral 2loop 2 1}). 
It remains to find the poles of $G(k_2,p_h)$. Due to the factor $\sigma(\Re\left\{k^0_1\right\})$, which implies the positivity of $\bar{k}^0_{1;4}$, only the last denominator in (\ref{integral II 0}) can be null, then $G(k_2,p_h)$ has only one pole at 
\be\label{equation pole k23}
\bar{k}^0_{1;4}= \sqrt{ \vec{k_1}^2 + m^2 - i \epsilon} \, ,
\ee
so that such pole is found solving  (\ref{equation pole k23}) for $\bar{k}_{2}^0$, which gives 
\be
\bar{k}_{2;3}^0= p^0 -\sqrt{\vec{k_1}^2+m^2  - i\epsilon}-\sqrt{(\vec{k}_1+\vec{k}_2-\vec{p})^2+m^2-i\epsilon} \, .
\ee 

\begin{figure}
	\begin{center}
		\includegraphics[height=6cm]{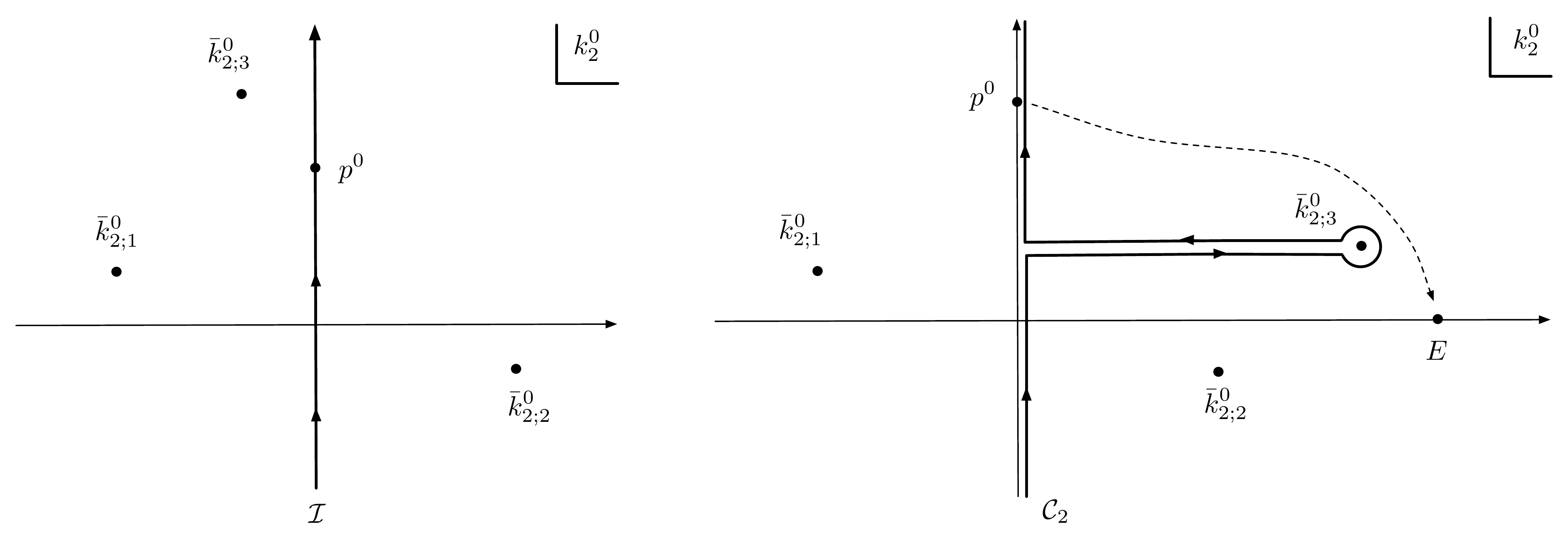}
		\caption{(Left) We plot the poles $\bar{k}^0_{2;1}, \bar{k}^0_{2;2},\bar{k}^0_{2;3}$ on the complex $k^0_2$ plane,  when $p^0$ is purely imaginary. (Right) We plot the same poles for $p^0$ real and positive. Since $\bar{k}^0_{2;1}$ and $\bar{k}^0_{2;2}$ do not depend on $p^0$, their positions do not change when $p^0 \rightarrow E$. On the contrary, $\bar{k}^0_{2;3}$ move to the right, and it passes through the imaginary axis $\mathcal{I}$ for some values of $\vec{k}_2$. We also plot the contour $\mathcal{C}_2$, which is obtained deforming $\mathcal{I}$ around $\bar{k}^0_{2;3}$.}
		\label{Fig7}
	\end{center}
\end{figure}

When the external energy $p^0$ is taken to be purely imaginary as it is in (\ref{amplitude2loop0a}), the poles $\bar{k}_{1;4}^0$ and $\bar{k}_{2;3}^0$ are on the left of the imaginary axis of the $k^0_1$ and $k^0_2$ complex planes, respectively (see Figs. \ref{Fig6} and  \ref{Fig7}). However, when $p^0 \rightarrow E\in \mathbb{R}^+_0$ both $\bar{k}_{1;4}^0$ and $\bar{k}_{2;3}^0$ pass through such imaginary axis and their real parts become simultaneously positive  for
\be\label{condition k2}
E > \Re\left\{\sqrt{\vec{k_1}^2+m^2  - i	\epsilon}+\sqrt{(\vec{k}_1+\vec{k}_2-\vec{p})^2+m^2-i\epsilon}\right\} \, ,
\ee 
(see  Fig.s \ref{Fig6} and \ref{Fig7}). Therefore, $\bar{k}_{2;3}^0$ contributes to the sum in (\ref{integral 2loop 2 1}) only when (\ref{condition k2}) is verified, and in that case $\Re\left\{\bar{k}_{2;3}^0\right\}>0$.
Therefore, the residue of $G(k_2,p_h)/(k_2^2-m^2+i\epsilon)$ at the pole $\bar{k}_{2;3}^0$ is:
\be
 && \hspace{-0.6cm}
{\rm Res} \left\{ \frac{G(k_2,p_h)}{(k_2^2-m^2+i\epsilon)},\bar{k}_{2;3}^0\right\}=
\int_{(\mathbb{R}^3)} \frac{i \, d^3 k_1}{(2\pi)^4}  \, 
 \left(   \frac{1}{k_2^2-m^2+i\epsilon}
\, 
\frac{(-2\pi i)\sigma\left(\Re\left\{k^0_1\right\}\right) }{2\sqrt{(\vec{p}-\vec{k}_1-\vec{k}_2)^2+m^2-i\epsilon}}\,\,
\frac{-\tilde{B}(k_1,k_2,p_h) }{2\sqrt{\vec{k_1}^2+m^2-i\epsilon}}\right)\Bigg|_{k_1^0=\bar{k}^0_{1;4}} ^{k_2^0=\bar{k}^0_{2;3}}  \, . \nonumber \\
&&\label{res G}
\ee
This expression can be used to evaluate (\ref{integral 2loop 2 1}), which reads
\be
\label{intG}
\hspace{-0.2cm}
\int_{(\mathcal{C}_2\times \mathbb{R}^3)} \!\! \frac{i \, d^4
	k_2}{(2\pi)^4}  \frac{ G(k_2,p_h) }{k_2^2 - m^2 + i \epsilon} \! = \!
\int_{(\mathcal{I}\times \mathbb{R}^3)} \!\!  \frac{i \, d^4
	k_2}{(2\pi)^4}  \frac{ G(k_2,p_h) }{k_2^2 - m^2 + i \epsilon}+ 
\int_{(\mathbb{R})^3} \!\!   
\frac{i \, d^3 k_2}{(2\pi)^4} \, (2 \pi i ) \sigma(\Re\left\{\bar{k}_{2;3}^0\right\})
{\rm Res} \left\{\frac{ G(k_2,p_h) }{k_2^2 - m^2 + i \epsilon},\bar{k}^0_{2;h}\right\} \! , 
\ee
where we have used the fact that $\bar{k}^0_{2;h}$ is the only pole contributing to the residues and we have introduced the term $\sigma(\Re\left\{\bar{k}_{2;3}^0\right\})$ since we know that the residue at $\bar{k}^0_{2;h}$ contributes only for $\bar{k}^0_{2;h}>0$. 

Since we are interested in the limit $\epsilon \rightarrow 0$ of the amplitude, we can neglect the $i \epsilon$ term in $\sqrt{(\vec{p}-\vec{k}_1-\vec{k}_2)^2+m^2-i\epsilon
}$ and $\sqrt{(\vec{p}-\vec{k}_1-\vec{k}_2)^2+m^2-i\epsilon
}$, while maintaining it in the propagator $k_2^2-m^2+i\epsilon$, so that  in the small $\epsilon$ limit we have
\be\label{resG}
&&\int_{(\mathbb{R})^3}\frac{i \, d^3 k_2}{(2\pi)^4} \, (2 \pi i ) \sigma(\Re\left\{\bar{k}_{2;3}^0\right\})
{\rm Res} \left\{\frac{ G(k_2,p_h) }{k_2^2 - m^2 + i \epsilon},\bar{k}^0_{2;h}\right\} = \quad \\
&&=\nonumber \int_{(\mathbb{R})^3}\frac{i \, d^3 k_2}{(2\pi)^4} \, (2 \pi i ) \sigma(\Re\left\{\bar{k}_{2;3}^0\right\})
\int_{(\mathbb{R}^3)} \frac{i \, d^3 k_1}{(2\pi)^4}  \, 
\left(   \frac{1}{k_2^2-m^2+i\epsilon}
\, 
\frac{(-2\pi i)\sigma\left(\Re\left\{k^0_1\right\}\right) }{2\sqrt{m^2+(\vec{k}_1+\vec{k}_2-\vec{p})^2}}\,\,
\frac{-\tilde{B}(k_1,k_2,p_h) }{2\sqrt{m^2 +\vec{k_1}^2  - i
		\epsilon}}\right)\Bigg|_{k_1^0=\bar{k}^0_{1;4}} ^{k_2^0=\bar{k}^0_{2;3}}  \\ \nonumber
&&= \int_{(\mathbb{R})^4}\frac{i \, d^4 k_2}{(2\pi)^4} \int_{(\mathbb{R}^4)} \frac{i \, d^4 k_1}{(2\pi)^4}  \, 
\frac{(-2\pi i)^2\,\sigma\left(k^0_1\right)\,\sigma\left(k^0_2\right)\,\sigma(p^0-k^0_1-k^0_2) \,\delta((p-k_1-k_2)^2-m^2)\,\delta(k_1^2-m^2)
	\,\tilde{B}(k_1,k_2,p_h) }{k_2^2 -m^2 + i \epsilon}\, , \nonumber 
\ee
where the last equality is obtained applying (\ref{delta complessa}) repeatedly. Using (\ref{integral II 0}) and (\ref{res G})-(\ref{intG}) we have

\be \nonumber
&& \hspace{-1cm}
\int_{(\mathcal{C}\times \mathbb{R}^3)} \frac{i \, d^4 k_2}{(2\pi)^4}  \frac{ G(k_2, p_h) }{k_2^2 - m^2 + i \epsilon}  \nonumber \\
&& \hspace{-1cm}
\nonumber
= \int_{(\mathcal{I}\times \mathbb{R}^3)} \frac{i \, d^4	k_2}{(2\pi)^4}  \frac{ 1}{k_2^2 - m^2 + i	\epsilon}\int_{(\mathbb{R}^3)} \,\frac{i \, d^3 k_1}{(2\pi)^4}  \! \left(
\frac{ (-2\pi i)\sigma\left(\Re\left\{k^0_1\right\}\right) }{2\sqrt{(\vec{k}_1+\vec{k}_2-\vec{p})^2+m^2-i\epsilon}}\,\,
\frac{\,\tilde{B}(k_1,k_2,p) }{k_1^2 - m^2 + i	\epsilon}\right)\Bigg|_{ k_1^0=\bar{k}^0_{1;4}}\\
&& \hspace{-1cm}
 \label{amplitude 2loop 8}
+ \int_{(\mathbb{R}^4)} \frac{i \, d^4 k_2}{(2\pi)^4}
\int_{(\mathbb{R}^4)} \frac{i \, d^4 k_1}{(2\pi)^4} \, \frac{\,(-2\pi i)^2
\sigma\left(k^0_1\right)\,\sigma\left(k^0_2\right)\,\sigma(p^0-k^0_1-k^0_2) \,\delta((p-k_1-k_2)^2-m^2)\,\delta(k_1^2-m^2)\tilde{B}(k_1,k_2,p) }{k_2^2 - m^2 +
i \epsilon} \, .
\ee
The first integral in (\ref{amplitude 2loop 8}) is real in the limit $\epsilon \rightarrow 0$,
since $k_2 \in \mathcal{I}\times \mathbb{R}^3$, $k^0_1=\bar{k}^0_{1;4}$, $\vec{k}_1 \in  \mathbb{R}^3$, thus $k_2^2 - m^2 \in \mathbb{R}$ while $k_1^2 - m^2\in \mathbb{R}$ for $\epsilon \rightarrow 0$, so that  it will not contribute to $\mathcal{M}-\mathcal{M}^*$  in such limit.

Finally, the analytic continuation (\ref{amplitude 2loop 1}) to real external energies of the complex amplitude (\ref{amplitude2loop0a}) is obtained substituting (\ref{amplitude 2loop 6}) and (\ref{amplitude 2loop 8}) in (\ref{amplitude 2loop 5}), obtaining

\be
&&
\hspace{-0.5cm}
\mathcal{M}(p_h,\epsilon) = - \frac{\lambda^2}{2}\left[\int_{(\mathcal{I}\times \mathbb{R}^3)} \frac{i \, d^4
	k_1}{(2\pi)^4}\frac{1}{k_1^2 - m^2 + i \epsilon}
\int_{(\mathcal{I}\times \mathbb{R}^3)} \frac{i \, d^4
	k_2}{(2\pi)^4} \,  \frac{1}{k_2^2 - m^2 + i
	\epsilon} \frac{\tilde{B}(k_1,k_2,p)}{(k_1+k_2-p)^2 - m^2 + i \epsilon} 
 \label{amplitude 2loop 9}
  \right. 
\\ 
&& \hspace{0.0cm} 
+\int_{(\mathcal{I}\times \mathbb{R}^3)} \frac{i \, d^4
	k_1}{(2\pi)^4}\frac{1}{k_1^2 - m^2 + i \epsilon}  
\int_{(\mathbb{R}^3)} \frac{i \, d^3 k_2}{(2\pi)^4} \, \left(-2\pi i\right) \,
\left(\frac{\sigma(\Re\left\{k^0_2\right\})}{2\sqrt{(\vec{p}-\vec{k}_1-\vec{k}_2)^2+m^2-i\epsilon
}} \,\,\frac{B(k,p_h)}{k_2^2 - m^2 + i	\epsilon}\right)\Bigg|_{k^0_2=\bar{k}^0_{2;4}}
\nonumber
\\
&& \hspace{0.0cm} 
+\int_{(\mathcal{I}\times \mathbb{R}^3)} \frac{i \, d^4	k_2}{(2\pi)^4}  \frac{ 1}{k_2^2 - m^2 + i	\epsilon}\int_{(\mathbb{R}^3)} \,\frac{i \, d^3 k_1}{(2\pi)^4}  \! \left(
\frac{ (-2\pi i)\sigma\left(\Re\left\{k^0_1\right\}\right) }{2\sqrt{(\vec{k}_1+\vec{k}_2-\vec{p})^2+m^2-i\epsilon}}\,\,
\frac{\,\tilde{B}(k_1,k_2,p) }{k_1^2 - m^2 + i	\epsilon}\right)\Bigg|_{ k_1^0=\bar{k}^0_{1;4}} \nonumber\\
&& \hspace{0.0cm}
\left. + \int_{(\mathbb{R}^4)} \frac{i \, d^4 k_2}{(2\pi)^4}
\int_{(\mathbb{R}^4)} \frac{i \, d^4 k_1}{(2\pi)^4} \, \frac{\,(-2\pi i)^2
	\sigma\left(k^0_1\right)\,\sigma\left(k^0_2\right)\,\sigma(p^0-k^0_1-k^0_2) \,\delta((p-k_1-k_2)^2-m^2)\,\delta(k_1^2-m^2)\tilde{B}(k_1,k_2,p) }{k_2^2 - m^2 +
	i \epsilon} \right] , \nonumber 
\ee

Therefore,  we can evaluate the limit $\epsilon \rightarrow 0$ of the imaginary part of the amplitude. We have already shown that the only term contributing is given by the last integral in (\ref{amplitude 2loop 9}), since all the other terms are real. Therefore we have 
\be
&&\hspace{-1cm}
\mathcal{M}(p_h,\epsilon)-\mathcal{M}(p_h,\epsilon)^* = - \frac{\lambda^2}{2}
\int_{(\mathbb{R}^4)} \frac{i \, d^4 k_1}{(2\pi)^4}
\int_{(\mathbb{R}^4)} \frac{i \, d^4 k_2}{(2\pi)^4} \, (-2\pi
i)^2\,\delta((p-k_1-k_2)^2-m^2)\,\delta(k_1^2-m^2)\,\times
\nonumber \\
&& \hspace{3cm}
\tilde{B}(k_1,k_2,p)  \,   \sigma\left(k^0_1\right)\,\sigma\left(k^0_2\right)\,\sigma(p^0-k^0_1-k^0_2) 
\left[\frac{1}{k_2^2 - m^2 + i \epsilon}-\frac{1}{k_2^2 - m^2 - i
\epsilon}\right] \,,
\ee
and using (\ref{delta}) we have

\be
&& \hspace{- .5cm}\nonumber
\lim_{\epsilon\rightarrow 0}\mathcal{M}(p_h,\epsilon)-\mathcal{M}(p_h,\epsilon)^* =- \frac{\lambda^2}{2} \!\! \int_{(\mathbb{R}^4)} \! \frac{i \, d^4
	k_1}{(2\pi)^4} \int_{(\mathbb{R}^4)} \! \frac{i \, d^4 k_2}{(2\pi)^4}
\, (-2\pi i)\,\sigma(p^0-k^0_1-k^0_2) \,\delta((p-k_1-k_2)^2-m^2)\, \\ \nonumber
\\ 
&& \hspace{5cm}
\times (-2\pi i)\,\sigma\left(k^0_1\right)\, \delta(k_1^2-m^2)\, (-2\pi i)\, \,\sigma\left(k^0_2\right)\, \delta(k_2^2-m^2) \,\tilde{B}(k_1,k_2,p) \,,
\ee
which confirms the validity of the Cutkosky rules for the two-loop process in Fig.\ref{Fig5}. Finally, in the special case in which the function $H(z)$ in  (\ref{vertex V definition}) is such that \mbox{$H(p^2= m^2)= 0$}, one has $\tilde{B}(k_1,k_2,p)=1$ when the internal momenta are on shell. Therefore, one has

\be
&& \hspace{-0.5cm}
\lim_{\epsilon\rightarrow 0}\mathcal{M}(E_h,\epsilon)-\mathcal{M}(E_h,\epsilon)^* = \\ \nonumber
&& \hspace{-0.5cm}
= - \frac{\lambda^2}{2} \!\! \int_{(\mathbb{R}^4)} \! \frac{i \, d^4
	k_1}{(2\pi)^4} \int_{(\mathbb{R}^4)} \! \frac{i \, d^4 k_2}{(2\pi)^4}\,
(-2\pi i)^3\,\sigma(p^0-k^0_1-k^0_2) \,\delta((p-k_1-k_2)^2-m^2)\,\sigma\left(k^0_1\right)\, \delta(k_1^2-m^2)\, \,\sigma\left(k^0_2\right)\, \delta(k_2^2-m^2) \, .
\nonumber
\ee

\section{Unitarity of the theory}\label{section unitarity}

In this section, we will complete the proof of the unitarity of the nonlocal theory (\ref{phin2}), showing that the imaginary part of the complex amplitudes  receive contributions only from Landau poles that correspond to normal thresholds, so that the unitary condition (\ref{unitarity condition M}) is fulfilled.

In the previous sections, we have shown how to compute the imaginary part of a scattering amplitude. Stating from (\ref{amplitude2bis}) or (\ref{Mba}) and applying repeatedly the residual formula  (\ref{formula integral 1}) according to (\ref{amplitude3}-\ref{amplitude3 integrated2}), one obtains the discontinuity in the imaginary part of the scattering amplitude as a sum of terms

\be \label{unitarity m 1}
&& 
\lim_{\epsilon\rightarrow 0}\mathcal{M}(E_h,\epsilon)-\mathcal{M}(E_h,\epsilon)^* = 
 \\ \nonumber 
 && \hspace{2cm} 
= - \frac{ \lambda^V  }{S_{\#} } \sum \int_{\Omega_1} \ldots
\int_{\Omega_L}\, \prod_{i=1}^L \frac{i
	\, d^4 k_i}{(2\pi)^4} \prod_{k=1}^N (-2\pi i)\,\delta(Q_k^2-m^2) \sigma(Q^0_k)
\prod_{j=1}^{I-N} \frac{1}{Q_j^2 - m^2 + i \epsilon} \,  
B(k_i,p_h)\, ,
\ee
where the $Q_k$ are the momenta corresponding to internal lines. Each term in the sum in (\ref{unitarity m 1}) corresponds to a cut diagram in which $N$ propagators are on shell while $I-N$ are not on shell. Moreover, for each term, the $i$-th integration region $\Omega_i$ can be 
$\mathbb{R}^4$ or $\mathcal{I}\times \mathbb{R}^3$, depending whether the corresponding momenta $k_i$ is contained in the propagator of one of the cut lines or not.

Let us consider the local version of the action  (\ref{phin2}), obtained assuming unitary form factor, i.e., $H(-\sigma \Box)\equiv 0$, that is,
\be
\mathcal{L}_{\phi} = \frac{1}{2} \partial_\mu \varphi \, \partial^\mu \varphi - \frac{1}{2} m^2 \varphi^2 - \lambda \sum_{n=4}^{N} \frac{c_n}{n !} \varphi^n \, .
\label{phin2 local}
\ee
Considering a generic amplitude $\mathcal{M}$, it is easy to show that, if a cut diagram does not contribute to   $\mathcal{M}-\mathcal{M}^*$ in the case of the local theory (\ref{phin2 local}), then the corresponding cut diagram in the nonlocal theory (\ref{phin2}) does not contribute to (\ref{unitarity m 1}).  The proof is obtained noting that for the local theory (\ref{phin2 local}) one has
\be \label{unitarity m 2}
 \lim_{\epsilon\rightarrow 0}\mathcal{M}(E_h,\epsilon)-\mathcal{M}(E_h,\epsilon)^*
= - \frac{ \lambda^V  }{S_{\#} } \sum \int_{\Omega_1} \ldots
\int_{\Omega_L}\, \prod_{i=1}^L \frac{i
	\, d^4 k_i}{(2\pi)^4} \prod_{k=1}^N (-2\pi i)\,\delta(Q_k^2-m^2) \sigma(Q^0_k)
\prod_{j=1}^{I-N} \frac{1}{Q_j^2 - m^2 + i \epsilon} \, ,
\ee
and (\ref{unitarity m 2}) contains the same terms with the same delta functions in (\ref{unitarity m 1}), with the only difference being that $B$ is replaced with one. This is due to the fact that $B(k_i,p_h) \neq 0$ by hypothesis; indeed' the nonlocality does not change the pole structure of the complex amplitudes, as discussed in Sec. \ref{Section Cutkosky rules}. Since the delta functions in (\ref{unitarity m 1}) and (\ref{unitarity m 2})  comes from the residues at the poles, it follows that each term in (\ref{unitarity m 1}) corresponds to a term in (\ref{unitarity m 2}) with the same delta functions.

Let us assume that the contribution of a specific cut diagram in (\ref{unitarity m 2}) is zero, i.e.,

\be 
&& \int_{\Omega_1} \ldots
\int_{\Omega_L}\, \prod_{i=1}^L \frac{i
	\, d^4 k_i}{(2\pi)^4} \prod_{k=1}^N (-2\pi i)\,\delta(Q_k^2-m^2) \sigma(Q^0_k)
\prod_{j=1}^{I-N} \frac{1}{Q_j^2 - m^2 + i \epsilon} \, = 0 \, .
\ee
When this happens, this quantity is null due to the fact that $\prod_{k=1}^N \,\delta(Q_k^2-m^2) \sigma(Q^0_k) = 0$, which implies that also  the corresponding term in (\ref{unitarity m 1}) is zero because it contains the same delta functions.  

Therefore, if a cut diagram does not contribute to (\ref{unitarity m 2}), it does not contribute to (\ref{unitarity m 1}).  Since the local theory (\ref{phin2 local}) is unitary, which can be demonstrated making use of the largest time equation \cite{largest time equation}, the sum (\ref{unitarity m 2}) does not contain contributions from cut diagrams corresponding to anomalous thresholds; indeed, the same will be true for (\ref{unitarity m 1}) in the case of our nonlocal theory. Furthermore, the unitarity of (\ref{phin2 local}) also implies that (\ref{unitarity m 2}) contains contributions from all the cut diagrams that appear on the rhs of (\ref{unitarity condition M}); indeed' the same will be true in the case of the nonlocal theory (\ref{phin2}). This, together with the Cutkosky rules, proves the unitarity of the nonlocal field theory (\ref{phin2}). 

We conclude this section noting that (\ref{unitarity m 1}) contains cut diagrams with at most $L+1$ cut lines, where $L$ is the number of loops in the diagram. This is easily understood if we think about how the delta functions arise. In fact, for each integration in (\ref{amplitude2bis}), one can apply  (\ref{formula integral 1}) obtaining a residue coming from the a pole of the integrand plus a term with no residue. Since each residue gives a delta function, in (\ref{amplitude2bis})  we have terms containing at most the product of $L$ delta functions. Indeed, (\ref{unitarity m 1}) contains  terms encompassing at most the product of $L+1$ delta functions, the last delta coming from the difference between the propagators, according to (\ref{delta}).  For instance, in the one loop diagram studied in Sec. \ref{Section one loop}, the only contribution to the imaginary part of the amplitudes comes from a diagram with two cut lines, while in the two-loop example of Sec. \ref{Section two loops}, one has only a cut  diagram containing three delta. This property of (\ref{unitarity m 1}) is very useful, since it makes possible to exclude immediately  the contribution of cut diagrams containing more than $L+1$ cut lines.

In what follows we will clarify what we have demonstrated above by means of two examples corresponding to the triangle and  box diagrams, which do have anomalous thresholds.

\subsection{Triangle diagram}\label{section triangle}

\begin{figure}
	\begin{center}
		\includegraphics[height=4cm]{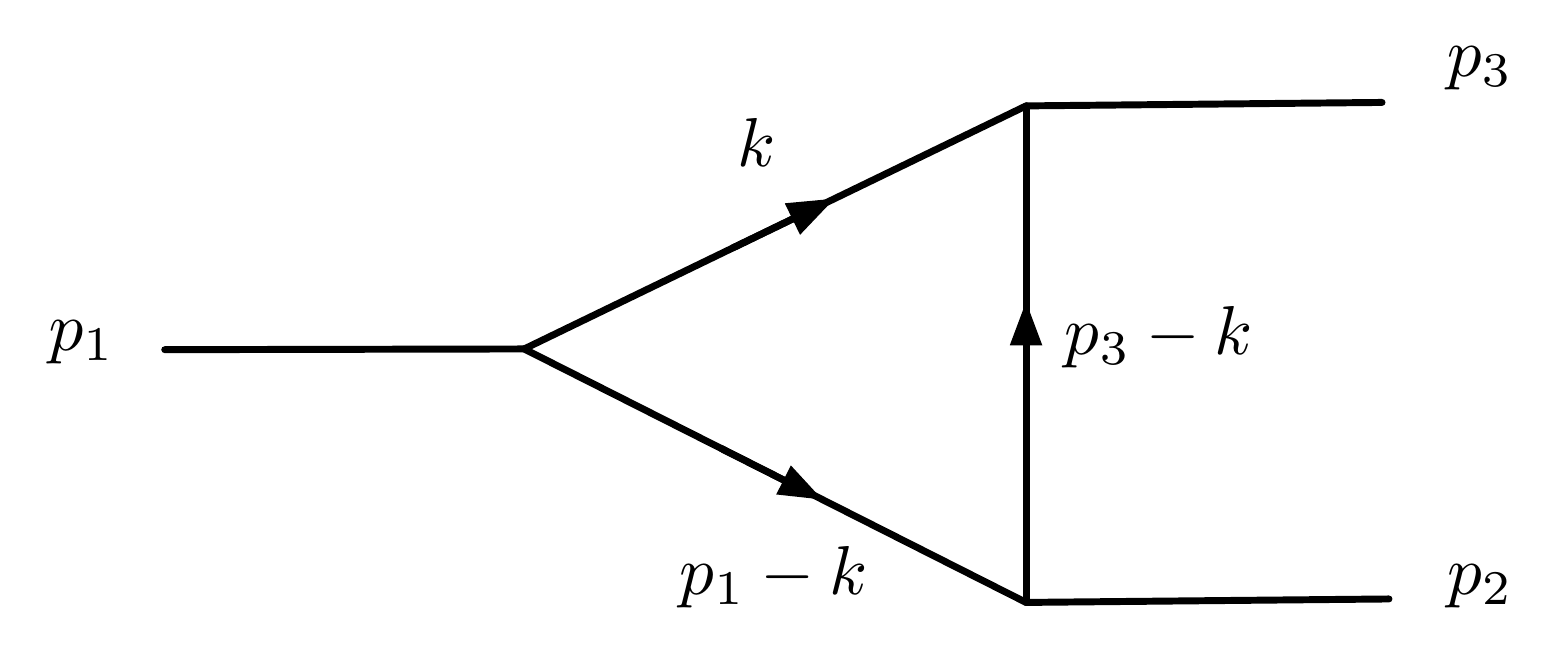}
		\caption{Triangle diagram. The kinematics allows for the three internal lines to go on shell at the same time. This gives a cut diagram that divides the triangle diagram in three parts when it is cut along the three on shell propagators. Such cut diagram corresponds to an anomalous threshold.}
		\label{Fig8} 
	\end{center}
\end{figure}

Let us consider the triangle diagrams in Fig.\ref{Fig8}. When the masses of the internal lines are chosen properly, the kinematics allows for the three internal lines to go on shell at the same time for some values of the external momenta. This situation corresponds to  an anomalous threshold since, cutting the lines corresponding to the three on shell propagators, the diagram is not divided in two parts. In such situation the question arises whether this anomalous threshold contributes to the lhs of (\ref{unitarity condition M}), since if this would be the case, such contribution could not be expressed as a product $ i \mathcal{M}^*_{c	b} \mathcal{M}_{c a}$ for some intermediate state $c$, and the theory would not be unitary.

Here we show that this is not the case, and (\ref{unitarity m 1}) does not contain contributions from such anomalous threshold. As we have discussed in the previous section, since the theory (\ref{phin2 local}) is unitary, we already know that the anomalous threshold of the triangle diagram does not contribute to (\ref{unitarity m 2}), therefore also (\ref{unitarity m 1}) does not contain contributions from this anomalous threshold.

Let us assume that $p_1$ is the momentum of the particle corresponding to the initial state, and $p_2$ and $p_3$ are the momenta of the particles in the final state. For the energy conservation, it will be $p_1 = p_2 + p_3$. The amplitude of the triangle diagram is 

\be\label{amplitude triangle 1 }
&&\mathcal{M}(p_h,\epsilon) = - \frac{ \lambda^2}{2} \, \int_{(\mathcal{C}\times
	\mathbb{R}^3)} \frac{i \, d^4 k}{(2\pi)^4}
\frac{1}{k^2 - m_1^2 + i \epsilon} \frac{1}{(k-p_1)^2 - m_2^2 + i \epsilon} \frac{B(k,p_1,p_2,p_3)}{(k-p_3)^2 - m_3^2 + i \epsilon}\, ,
\ee
where $B(k,p_1,p_2,p_3) \equiv  \mathcal{V}(p_1,k,p-k) \, \mathcal{V}(p_1-k,p_2,p_3-k) \,\mathcal{V}(p_3-k,p_3,k)$. Since we want this amplitude to be nonzero, we assume $p_1^2 = M^2 > p_2^2 + p_3^2$. We also assume that the mass of the internal propagators are different, so that the three propagators can go on shell together.  For instance, this happens in the case $m_1 = m_2 = m$, and $p^2_2 = p^2_3 = p^2 > m^2 + m_3^2$, when one has an anomalous threshold at 

\be
p_1^2 = 4 m^2 - \frac{(p^2-m^2-m_3^2)^2}{m_3^2} ,
\ee
see \cite{itzykson zuber} for details. Anomalous thresholds occur also in other configurations, e.g., in the case $m_1=m_3 =m$, $m_2 = 2 m$, $p_2^2 = m^2$, $p_3^2 = 4 m^2$ for $p_1^2 = 9 m^2$.

Even before starting our calculation, we already know that the anomalous threshold corresponding to the cut diagram with the three internal lines on shell cannot contribute to the imaginary part of $\mathcal{M}$. In fact, we have already shown that $\mathcal{M}-\mathcal{M}^*$ contains terms encompassing at most the product of $L+1$  delta functions, namely at most two cut lines in the case of the one-loop triangle diagram under consideration. Therefore, the anomalous threshold corresponding to three cut lines does not contribute to $\mathcal{M}-\mathcal{M}^*$. An explicit calculation will show that we have contributions from the same cut diagrams that contribute in the case of the local theory (\ref{phin2 local}), as expected. This is due to the fact that $B(k,p_1,p_2,p_3)$ has no zeros or poles, thus the local and nonlocal fields (\ref{phin2}) and (\ref{phin2 local}) have the same singularity structure.

\begin{figure}
	\begin{center}
		\includegraphics[height=6cm]{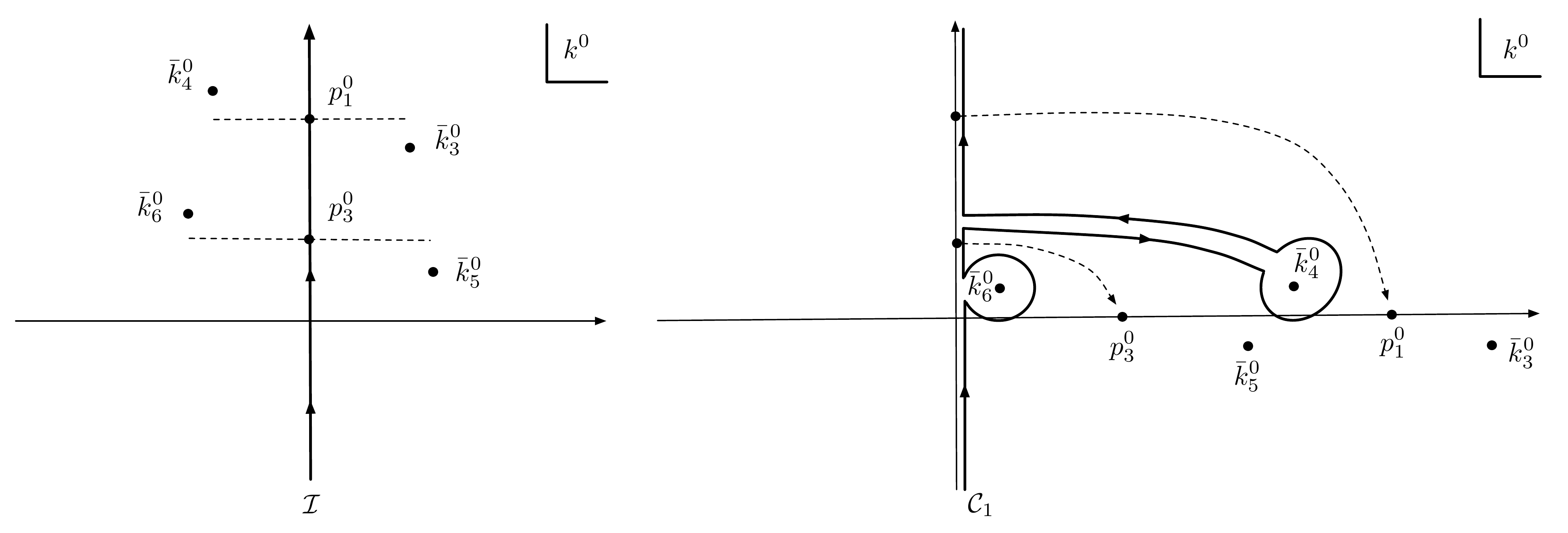}
		\caption{(Left) We plot the poles $\bar{k}^0_{3}, \bar{k}^0_{4}, \bar{k}^0_{5}, \bar{k}^0_{6}$ on the complex $k^0$ plane,  when the external energies $p^0_1$ and $p^0_3$ are purely imaginary. (Right) We plot the same poles when the external energies are moved to their physical real and positive values. The poles $\bar{k}^0_{4}$ and $\bar{k}^0_{6}$ pass through the imaginary axis for some values of the loop momenta $k$, giving the residues in (\ref{amplitude triangle 2 }).}
		\label{Fig9} 
	\end{center}
\end{figure}

We know that the integration contour $(\mathcal{C}\times \mathbb{R}^3)$ in (\ref{amplitude triangle 1 }) is obtained deforming the imaginary axis $\mathcal{I}$  around the poles of the propagators in (\ref{amplitude triangle 1 }) that pass through it when the energies $p^0_1, p^0_2, p^0_3$ become real. The poles of the first propagator are
\bea
\bar{k}^0_{1,2}
= \pm \sqrt{\vec{k}^2 + m_1^2 - i \epsilon}\, . 
\eea
Such poles do not depend on the
external energies, and  remain always far from the imaginary
axis. Indeed, $\bar{k}^0_{1,2}$ do not pinch $\mathcal{I}$ and do not contribute to the sum of
residues in (\ref{formula integral 1}). The poles of the second and third propagators are 
\bea
\bar{k}^0_{3,4} = p^0_1 \pm
\sqrt{(\vec{k}-\vec{p}_1)^2+m_2^2-i\epsilon } \, , \quad \bar{k}^0_{5,6} = p^0_3 \pm
\sqrt{(\vec{k}-\vec{p}_3)^2+m_3^2-i\epsilon }. 
\eea
When the external energies are purely imaginary,  $\bar{k}^0_{3}$ and $\bar{k}^0_{5}$ are at the right and
$\bar{k}^0_{4}$ and $\bar{k}^0_{6}$ are at the left of the imaginary axis of the $k^0$
plane. When the external energies are moved to their physical real and positive values, such poles move to the right, and it happens that the poles $\bar{k}^0_{4}$ and $\bar{k}^0_{6}$ pass through the imaginary axis for some values of the loop momenta $k$, see Fig.\ref{Fig9}. Therefore, the integration contour $\mathcal{C}$ is obtained deforming the imaginary axis $\mathcal{I}$ around the poles
$\bar{k}^0_{4}$ and $\bar{k}^0_{6}$ when such poles pass through
$\mathcal{I}$, i.e.,  when $\Re\left\{\bar{k}^0_{4}\right\} > 0$ and $\Re\left\{\bar{k}^0_{6}\right\} > 0$ respectively.

According to (\ref{formula integral 1}), we can express (\ref{amplitude triangle 1 }) as 
\be
&&\label{amplitude triangle 2 }
\hspace{-1.5cm}
\mathcal{M}(p_h,\epsilon)  = - \frac{ \lambda^2}{2}    \left[ \int_{(\mathcal{I}\times
	\mathbb{R}^3)} \frac{i \, d^4 k}{(2\pi)^4}
\frac{1}{k^2 - m_1^2 + i \epsilon} \frac{B(k,p_1,p_2,p_3)}{(k-p_1)^2 - m_2^2 + i \epsilon} \frac{1}{(k-p_3)^2 - m_3^2 + i \epsilon} \right. \\
&&  \hspace{0.4cm}
 +
\int_{(\mathbb{R}^3)} \frac{i \, d^3 k}{(2\pi)^4} \, 2\pi i \,
\sigma(\Re\left\{\bar{k}^0_{4}\right\}) \, {\rm Res} \left\{\frac{B(k,p_1,p_2,p_3)}{k^2 - m_1^2 + i \epsilon} \frac{1}{(k-p_1)^2 - m_2^2 + i \epsilon} \frac{1}{(k-p_3)^2 - m_3^2 + i \epsilon},\bar{k}^0_{4}\right\} \nonumber
\\
&&\hspace{0.4cm}
 \left.  +
\int_{(\mathbb{R}^3)} \frac{i \, d^3 k}{(2\pi)^4} \, 2\pi i \,
\sigma(\Re\left\{\bar{k}^0_{6}\right\}) \, {\rm Res} \left\{\frac{B(k,p_1,p_2,p_3)}{k^2 - m_1^2 + i \epsilon} \frac{1}{(k-p_1)^2 - m_2^2 + i \epsilon} \frac{1}{(k-p_3)^2 - m_3^2 + i \epsilon},\bar{k}^0_{6}\right\} \right] . \nonumber
\ee
The residues in (\ref{amplitude triangle 2 }) are easily calculated, giving 
\be
&&{\rm Res} \left\{\frac{1}{k^2 - m_1^2 + i \epsilon} \frac{1}{(k-p_1)^2 - m_2^2 + i \epsilon} \frac{B(k,p_1,p_2,p_3)}{(k-p_3)^2 - m_3^2 + i \epsilon},\bar{k}^0_{4}\right\} \nonumber\\
&& \hspace{1cm}
= \left(
- \frac{B(k,p_1,p_2,p_3)}{k^2 - m_1^2 + i \epsilon} \frac{1}{2 \sqrt{(\vec{k}-\vec{p}_1)^2 + m_2^2  -i \epsilon}
} \frac{1}{(k-p_3)^2 - m_3^2 + i \epsilon} \right)\Bigg|_{k^0 = \bar{k}^0_{4}} ,
\ee
and 
\be
&&{\rm Res} \left\{\frac{1}{k^2 - m^2 + i \epsilon} \frac{1}{(k-p_1)^2 - m^2 + i \epsilon} \frac{B(k,p_1,p_2,p_3)}{(k-p_3)^2 - m^2 + i \epsilon},\bar{k}^0_{6}\right\} \nonumber\\
&& \hspace{1cm}
= \left(
- \frac{B(k,p_1,p_2,p_3)}{k^2 - m^2 + i \epsilon} \frac{1}{2 \sqrt{(\vec{k}-\vec{p}_3)^2 + m^2 - i \epsilon}
} \frac{1}{(k-p_1)^2 - m^2 + i \epsilon} \right)\Bigg|_{k^0 = \bar{k}^0_{6}},
\ee
so that the amplitude (\ref{amplitude triangle 2 }) becomes

\be
&&\label{amplitude triangle 3 }
\mathcal{M}(p_h,\epsilon)  = - \frac{ \lambda^2}{2}    \left[ \int_{(\mathcal{I}\times
	\mathbb{R}^3)} \frac{i \, d^4 k}{(2\pi)^4}
\frac{1}{k^2 - m_1^2 + i \epsilon} \frac{B(k,p_1,p_2,p_3)}{(k-p_1)^2 - m_2^2 + i \epsilon} \frac{1}{(k-p_3)^2 - m_3^2 + i \epsilon} \right. \\
&&   +
\int_{(\mathbb{R}^3)} \frac{i \, d^3 k}{(2\pi)^4} \, (-2\pi i) \,
\sigma(\Re\left\{\bar{k}^0_{4}\right\}) \,  \nonumber \left( \frac{B(k,p_1,p_2,p_3)}{k^2 - m_1^2 + i \epsilon} \frac{1}{2 \sqrt{(\vec{k}-\vec{p}_1)^2 + m_2^2  -i \epsilon}
} \frac{1}{(k-p_3)^2 - m_3^2 + i \epsilon} \right)\Bigg|_{k^0 = \bar{k}^0_{4}}
\\
&&  +
\int_{(\mathbb{R}^3)} \frac{i \, d^3 k}{(2\pi)^4} \, (-2\pi i) \,
\sigma(\Re\left\{\bar{k}^0_{6}\right\}) \,\left(
- \frac{B(k,p_1,p_2,p_3)}{k^2 - m_1^2 + i \epsilon} \frac{1}{2 \sqrt{(\vec{k}-\vec{p}_3)^2 + m_3^2 - i \epsilon}
} \frac{1}{(k-p_1)^2 - m_2^2 + i \epsilon} \right)\Bigg|_{k^0 = \bar{k}^0_{6}} . \nonumber
\ee
Finally, (\ref{amplitude triangle 3 }) can be recast by means of (\ref{delta complessa}) as

\be
&& \hspace{-2cm}
\label{amplitude triangle 4 }
\mathcal{M}(p_h,\epsilon)  = - \frac{ \lambda^2}{2}    \left[ \int_{(\mathcal{I}\times
	\mathbb{R}^3)} \frac{i \, d^4 k}{(2\pi)^4}
\frac{1}{k^2 - m_1^2 + i \epsilon} \frac{B(k,p_1,p_2,p_3)}{(k-p_1)^2 - m_2^2 + i \epsilon} \frac{1}{(k-p_3)^2 - m_3^2 + i \epsilon} \right. \\
&&   \hspace{0.5cm}
+
\int_{(\mathbb{R}^4)} \frac{i \, d^4 k}{(2\pi)^4} \, (-2\pi i) \,
\sigma({k}^0) \, \sigma(p^0_1-k^0) \, \nonumber  \frac{B(k,p_1,p_2,p_3)}{k^2 - m_1^2 + i \epsilon}  \frac{\delta((k-p_1)^2 - m^2_2)}{(k-p_3)^2 - m_3^2 + i \epsilon} 
\\
&&  \hspace{0.5cm} \left.
+
\int_{(\mathbb{R}^4)} \frac{i \, d^4 k}{(2\pi)^4} \, (-2\pi i) \,
\sigma({k}^0) \, \sigma(p^0_3-k^0) \,\frac{B(k,p_1,p_2,p_3)}{k^2 - m_1^2 + i \epsilon}  \frac{\delta((k-p_3)^2 - m^2)}{(k-p_1)^2 - m_2^2 + i \epsilon} \right]. \nonumber
\ee
It is easy to  see that the first integral in (\ref{amplitude triangle 4 }) is real in the limit $\epsilon \rightarrow 0$, 
since $k \in \mathcal{I}\times \mathbb{R}^3$, indeed the three denominators in the integrand are never zero, and they become real in such a limit. Therefore, such integral does not contribute to the imaginary part of the complex amplitude, and one has
\bea\label{amplitude triangle 5}
&& \hspace{-.3cm}
\mathcal{M}(p_h,\epsilon)  -\mathcal{M}(p_h,\epsilon) ^* = \nonumber\\
&& \hspace{-.3cm}- \frac{ \lambda^2}{2}\int_{(\mathbb{R}^4)} \frac{i \, d^4 k}{(2\pi)^4} \, (-2\pi i)^2\, \left[\sigma({k}^0) \, \sigma(p^0_1-k^0)\left(\frac{\delta(k^2 - m_1^2)}{(k-p_3)^2 - m_3^2 + i \epsilon} + \frac{\delta((k-p_3)^2- m_3^2)}{k^2  - m_1^2 + i \epsilon} \right) \delta((p_1-k)^2- m_2^2)\right. \nonumber\\
&&\hspace{-.3cm}\left.+\sigma({k}^0) \, \sigma(p^0_3-k^0)\left(\frac{\delta(k^2 - m_1^2)}{(k-p_1)^2 - m_2^2 + i \epsilon} + \frac{\delta((k-p_1)^2- m_2^2)}{k^2  - m_1^2 + i \epsilon} \right) \delta((p_3-k)^2- m_3^2) \right] \,  B(k,p_1,p_2,p_3).
\eea
Finally, we note that the kinematics implies that, since $p_1^2 > p_2^2 + p_3^2$, the only propagators that can go on shell together are those involving the momenta $k$ and $p_1-k$, so that one  has
\bea\label{amplitude triangle 6}
&& \hspace{-1.5cm}
\mathcal{M}(p_h,\epsilon)  -\mathcal{M}(p_h,\epsilon) ^* =- \frac{ \lambda^2}{2} \int_{(\mathbb{R}^4)} \frac{i \, d^4 k}{(2\pi)^4} \, (-2\pi i)^2 B(k,p_1,p_2,p_3) \frac{\sigma({k}^0) \, \sigma(p^0_1-k^0)}{(k-p_3)^2 - m_3^2 + i \epsilon} \,\delta(k^2 - m_1^2)\,\delta((p_1-k)^2 - m_2^2) .
\eea

\begin{figure}
	\begin{center}
		\includegraphics[height=4cm]{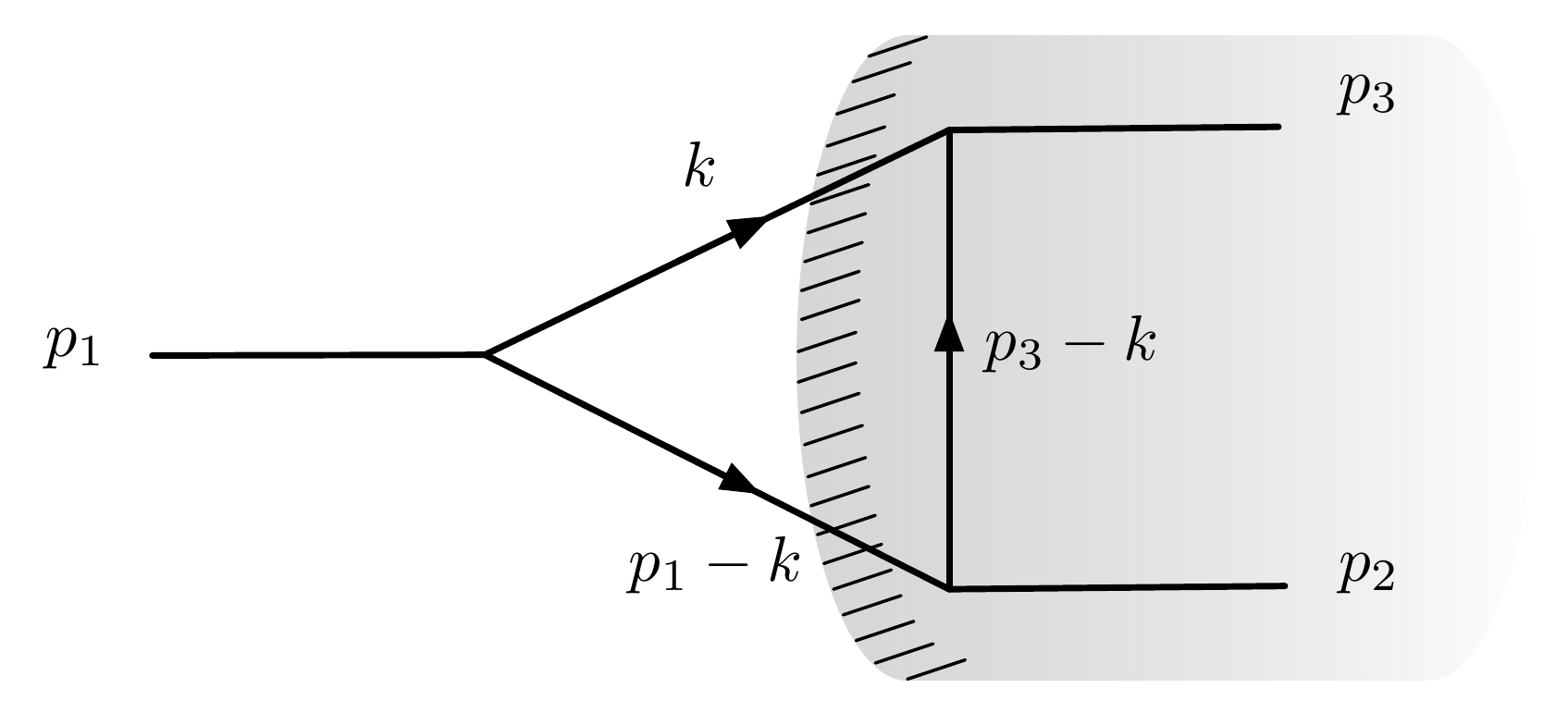}
		\caption{We here plot the only cut diagram that contributes to $\mathcal{M}  -\mathcal{M}^*$. In fact, due to the kinematic condition  $p_1^2 > p_2^2 + p_3^2$, the only couple of propagators that can go on shell together is that involving the momenta $k$ and $p_1-k$. According to the standard notation, the energy flows from the unshadowed to the shadowed region.}
		\label{Fig11} 
	\end{center}
\end{figure}

From (\ref{amplitude triangle 6}) we see that, due to the kinematics, only one of the cut diagrams with two cut lines, more precisely, the one shown in Fig. \ref{Fig11}, contributes to $\mathcal{M}  -\mathcal{M}^*$, while the anomalous threshold of the triangle diagram does not contribute to $\mathcal{M}  -\mathcal{M}^*$. This is exactly what we were expecting, because, apart from the nonlocal term $B(k,p_1,p_2,p_3)$, (\ref{unitarity m 1}) is the same as in local theory. Therefore, it cannot contain other diagrams than those of the the local theory. 
This example confirms what we stated in the Sec. \ref{section unitarity} on the absence of contributions from anomalous thresholds and on the unitarity of the nonlocal theory.

\subsection{Box diagram}\label{section Box}

\begin{figure}
	\begin{center}
		\includegraphics[height=4cm]{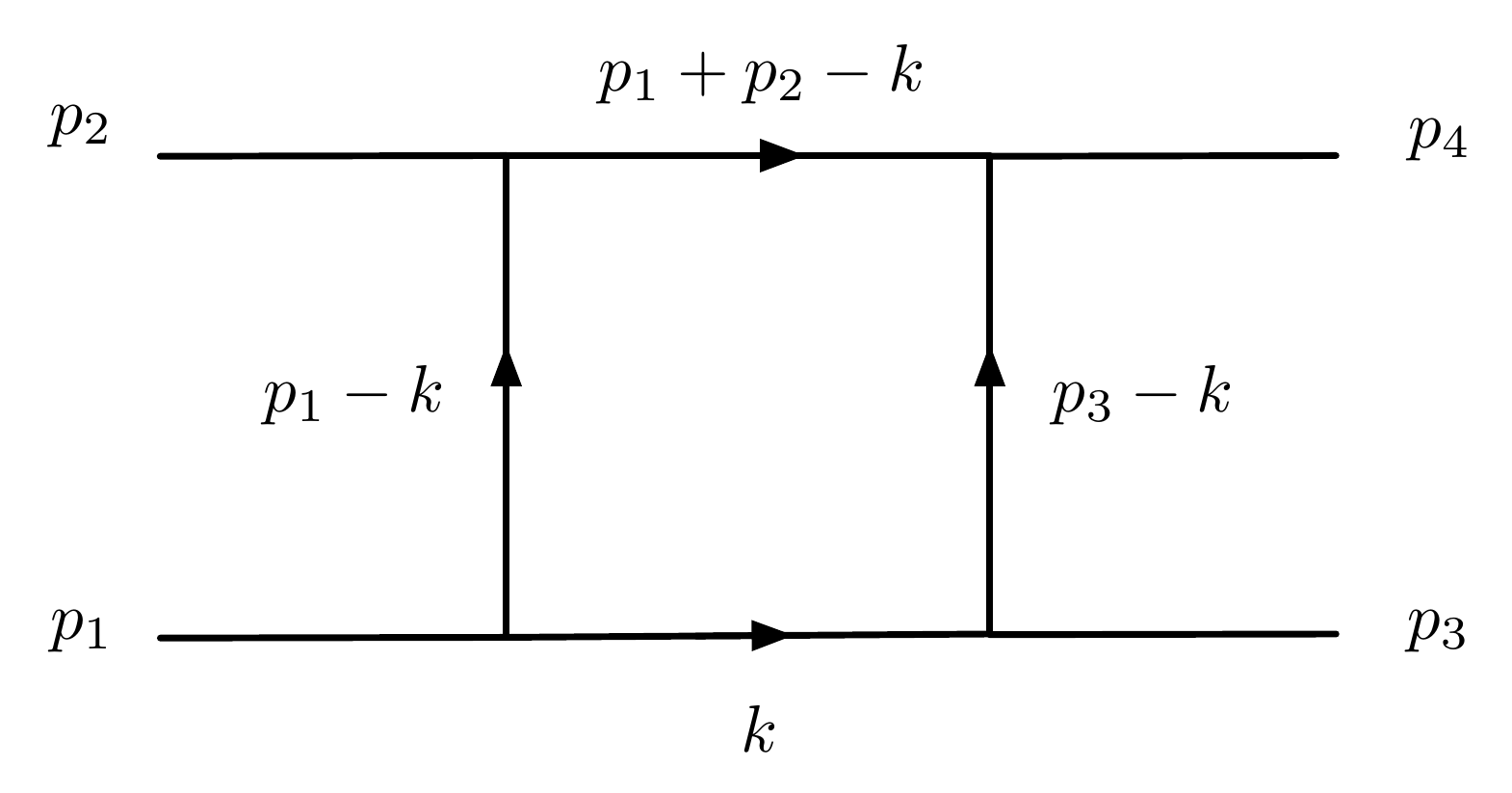}
		\caption{Box diagram. The momenta $p_1$ and $p_2$ correspond to the initial state, while $p_3$ and $p_4$ are the momenta of the final state. Each internal propagator has a different mass, and  depending on the choice of such masses, one can have anomalous thresholds with three or four internal momenta on shell.}
		\label{Fig12} 
	\end{center}
\end{figure}

In this section, we study the amplitude of the box diagram in Fig.\ref{Fig12}. 
We assume that $p_1$ and $p_2$ are  the momenta in the initial state, and $p_3$ and $p_4$ are the momenta in the final state. In a theory with cubic interactions, such momenta corresponds to the momenta of single particles; indeed, they must be on shell. On the contrary, in the case of a quartic interaction, they correspond to the momenta of a couple of particles; indeed, they are not on shell. In both cases, the energy conservation implies $p_1 + p_2 = p_3 + p_3$. Here we treat the external momenta as independent variables that do not need to be   on shell. 
The amplitude of the box diagram is
\be &&
\label{amplitude box 1 }
\mathcal{M}(p_h,\epsilon) = - \frac{ \lambda^2}{2} \, \int_{(\mathcal{C}\times	\mathbb{R}^3)} \frac{i \, d^4 k}{(2\pi)^4}
\frac{1}{k^2 - m_1^2 + i \epsilon} \quad \frac{1}{(p_1-k)^2 - m^2_2 + i \epsilon} \quad\frac{1}{(p_1+p_2-k)^2 - m^2_3 + i \epsilon}\times \\
&& \hspace{4.7cm}
\nonumber  \frac{1}{(p_3-k)^2 - m^2_4 + i \epsilon}\, B(k,p_1,p_2,p_3,p_4)\, ,
\ee
where 
\be 
B(k,p_1,p_2,p_3) \equiv  \mathcal{V}(p_1,k,p_1-k) \, \mathcal{V}(p_1-k,p_2,p_1+p_2-k)\,\mathcal{V}(p_1+p_2-k, p_4,p_3-k)\,\mathcal{V}(p_3-k,p_3,k).
\ee 
We allow the masses of the internal propagators to be different, since we want to consider the leading anomalous threshold in which the four propagators can go on shell at the same time. For instance, if $m_1 = 3m$ and $m_2=m_3=m_4= m$, the four propagators can go on shell at the same time. Instead, if all the masses are equal, say $m_1 = m_2=m_3=m_4= m$, only three propagators can go on shell at the same time. 

Based on the properties of (\ref{unitarity m 1}) stated in Sec. \ref{section unitarity}, the imaginary part of  the complex amplitude  contains contributions coming from cut diagrams with at most $L+1$ cut lines, where $L$ is the number of loops. Therefore, in our one-loop box diagram the contributions to (\ref{amplitude box 1 }) come from diagrams with at most two cut lines, which automatically exclude the anomalous thresholds mentioned above. Since cut diagrams with only one propagator on shell do not give any singularity, because one needs at least two propagators on shell to constrain the integration contour between two poles,   we remain with contributions coming only from cut diagrams with two cut lines. Among all the diagrams with two cut lines, the kinematics selects those diagrams  allowed by the energy conservation.

Using the usual procedure given  by the iterative relations  (\ref{amplitude3 integrated}) and (\ref{amplitude3 integrated2}) and  the energy conservation $p_1 + p_2 = p_3 + p_4$, one has
\bea
\label{amplitude box 2}
&& \hspace{-1.5cm}
\mathcal{M}(p_h,\epsilon)  -\mathcal{M}(p_h,\epsilon) ^* =- \frac{ \lambda^2}{2} \int_{(\mathbb{R}^4)} \sigma(k^0) \frac{i \, d^4 k}{(2\pi)^4} \, (-2\pi i)^2 \delta(k^2 - m_1^2)\,\sigma(p^0_1+p^0_2-k^0)\delta((p_1+p_2-k)^2 - m_3^2) \times \\
&& \hspace{3cm}
 \nonumber \frac{1}{(p_1-k)^2 - m^2_2 + i \epsilon} \,  \frac{1}{(p_3-k)^2 - m^2_4 + i \epsilon}\, B(k,p_1,p_2,p_3,p_4) \, ,
\eea
which corresponds to the same cut diagram that contributes in the local case. Therefore, the unitarity of the box diagram is guaranteed.

\section{Nonlocal quantum gauge theories and gravity}\label{section gauge gravity}
The Cutkosky rules and the unitarity condition (\ref{unitarityM}) have been derived for a scalar field, but it is quite easy to generalize them  to nonlocal gauge theories (NLGT) or nonlocal gravity (NLG). These theories are invariant under gauge or general coordinate transformations that must be treated properly to proof unitarity. In order to prove unitarity, we need to demonstrate the cancellation of the unphysical cuts, which is actually not so different from the local gauge or gravity case. 
In the scalar field theory, the tensorial structure is trivial, but in NLGT and NLG some of the components of the propagators seem to violate unitarity. However, this is not the  case because of the presence of the Faddeev-Popov (FP) ghosts  that provide the right cancellations in the amplitude to allow only physical states (physical gauge-invariant polarizations) to go on shell. 
What we need to use to prove unitarity are the Ward identities that have been derived for a general field theory in Sec. (2.7) of \cite{Shapirobook} and more recently in \cite{Lavrov:2019nuz}. Indeed, the proof of the cancellation of unphysical cuts relies only on the gauge or diffeomorphism (or their quantum BRST version) invariance of the action, and 
it is usually stated at the formal level of the path integral independence of the gauge-fixing term, which, of course, holds here as well.

For NLGT and NLG, we have to sum the contributions coming from gauge bosons and the FP ghosts, but all the amplitudes are analytically continued regardless of the particle species, and the Cutkowsky rules are the same derived for the real scalar field up to an overall tensorial structure.

\section{CONCLUSIONS} \label{section conclusions}
We have proved the Cutkosky rules and  the perturbative unitarity for Euclidean nonlocal scalar field and discussed their generalization to gauge and gravitational theories. In short, any scattering amplitude can be computed in Euclidean signature in which all external (and internal) energies are taken purely imaginary. Afterwards, the amplitudes can be analytically continued  making the external energies real. The latter operation actually corresponds to integrate along sophisticated paths in the complex plane of the energies circulating in the loops' integrals. It turns out that the analytically continued amplitudes satisfy the Cutkosky rules, and that anomalous thresholds do not contribute to (\ref{unitarityM}), so that the perturbative unitarity of the diagrams is proved.  Moreover, we argued that the BRST invariance and the Ward identities imply  that all the unphysical cuts present in gauge and gravitational theories cancel with similar contributions coming from the Faddeev-Popov ghosts, and only the physical polarizations can go on shell.

We stress one more time  that the fact that the Cutkosky rules are still valid for nonlocal field theories is not surprising. In fact, the Cutkosky result \cite{cutkosky} is derived only on the basis of the analysis of the poles of the propagators, and the form of the vertices, which can be polynomial or weakly nonlocal, does not play any role in such derivation. In nonlocal theories, the singularities of the amplitudes are still determined  by the Landau equations \cite{landau}, and, therefore, the diagrams in local and nonlocal theories have the same singularities and branch cuts for the same values of the external momenta, which corresponds to the same thresholds. 
Finally, (\ref{unitarityM}) receives contributions from the same cut diagrams of the local theory (\ref{phin2 local}); indeed, contributions from anomalous thresholds are absent, and the nonlocal theory is unitary.



\begin{thebibliography}{99}

\bibitem{Krasnikov}
  N.~V.~Krasnikov,
  ``Nonlocal Gauge Theories,''
  Theor.\ Math.\ Phys.\  {\bf 73}, 1184 (1987)
  [Teor.\ Mat.\ Fiz.\  {\bf 73}, 235 (1987)].

\bibitem{kuzmin}
  Y.~V.~Kuz'min,
  ``The Convergent Nonlocal Gravitation (in Russian),''
  Sov.\ J.\ Nucl.\ Phys.\  {\bf 50}, 1011 (1989)
  [Yad.\ Fiz.\  {\bf 50}, 1630 (1989)].




\bibitem{shapiro asorey}  
M. Asorey, J.L. Lope, I.L. Shapiro, \textit{Some remarks on high derivative quantum gravity}, Int.J.Mod.Phys. A12 (1997) 5711-5734.

\bibitem{modesto}
  L.~Modesto,
  ``super-renormalizable Quantum Gravity,''
  Phys. \ Rev. \ D {\bf 86}, 044005 (2012)
  [arXiv:1107.2403 [hep-th]].


\bibitem{Modesto:2012ys} 
  L.~Modesto,
  ``superrenormalizable Multidimensional Quantum Gravity,''
  Astron.\ Rev.\  {\bf 8}, no. 2, 4 (2013)
  [arXiv:1202.3151 [hep-th]].


   \bibitem{modestoLeslaw}
  L.~Modesto and L.~Rachwal,
  ``Super-renormalizable and finite gravitational theories,''
  Nucl.\ Phys.\ B {\bf 889}, 228 (2014)
  [arXiv:1407.8036 [hep-th]].

\bibitem{Modesto:2017sdr} 
  L.~Modesto and L.~Rachwal,
  ``Nonlocal quantum gravity: A review,''
  Int.\ J.\ Mod.\ Phys.\ D {\bf 26}, no. 11, 1730020 (2017).

\bibitem{reflection positivity} 
M. Christodoulou, L. Modesto, \textit{ Reflection positivity in nonlocal gravity}, arXiv:1803.08843 [hep-th]. M. Asorey, L. Rachwal, I. L. Shapiro, \textit{Unitary Issues in Some Higher Derivative Field Theories 
}, Galaxies 6 (2018) no.1, 23.




\bibitem{peskin} M. E. Peskin, D. V. Schroeder, ``An Introduction To Quantum Field Theory", Avalon Publishing, 1995.



\bibitem{itzykson zuber} C. Itzykson, J. B. Zuber, ``Quantum Field Theory",
Dover publications Inc. 2006, ISBN-10: 0486445682. ISBN-13:
978-0486445687.


\bibitem{antoniadis} I. Antoniadis, E. T. Tomboulis, \textit{Gauge invariance and unitarity in higher-derivative quantum gravity}, Phys. Rev. D {\bf 33}, 2756 (1986).



\bibitem{Modesto:2015lna} 
  L.~Modesto and L.~Rachwal,
  ``Universally finite gravitational and gauge theories,''
  Nucl.\ Phys.\ B {\bf 900}, 147 (2015)
  [arXiv:1503.00261 [hep-th]].


\bibitem{piva} 
  L.~Modesto, M.~Piva and L.~Rachwal,
  ``Finite quantum gauge theories,''
  Phys.\ Rev.\ D {\bf 94}, no. 2, 025021 (2016)
  [arXiv:1506.06227 [hep-th]].




\bibitem{landau} L. D. Landau, ``On analytic properties of vertex parts in quantum field theory", Nuclear Phys. {\bf 13}, 181 (1959)



\bibitem{cutkosky} R. E. Cutkosky, ``Singularities and Discontinuities of Feynman Amplitudes", Journal of Mathematical Physics {\bf 1}, 429 (1960).


\bibitem{taylor} J. C. Taylor, ``Analytic Properties of Perturbation Expansions", Phys. Rev. {\bf 117}, 261 (1960).

\bibitem{Shapirobook} I. L. Buchbinder, S. D. Odintsov, I. L. Shapiro, 
  ``Effective action in quantum gravity", IOP Publishing Ltd 1992. 






\bibitem{largest time equation} M. J. G. Veltman, Unitarity and causality in a renormalizable field theory with unstable particles, Physica {\bf 29}, 186 (1963); G. ’t Hooft, M. J. G. Veltman, “Diagrammar,” NATO Sci. Ser. B {\bf 4}, 177 (1974).


\bibitem{Lavrov:2019nuz} 
  P.~M.~Lavrov and I.~L.~Shapiro,
  arXiv:1902.04687 [hep-th].







\end{thebibliography}
\end{document}